\documentclass[pra,aps,reprint,superbib,superscriptaddress,twocolumn]{revtex4-1}
\usepackage{amsmath,bm,bbm}
\usepackage{amstext}
\usepackage{epsfig}
\usepackage{xcolor}
\usepackage{subfig}
\usepackage{graphicx}
\usepackage{multirow}
\usepackage{array}
\usepackage{tikz,pgfplots}
\usepackage{MnSymbol}
\usepackage{dsfont}
\usepackage{bbold}
\usepackage{physics}
\usepackage{mathtools}
\usepackage{hyperref}
\usepackage{bm}

\usepackage{algorithm}
\usepackage{algorithmic}

\usepackage{ulem}
\hypersetup{
    colorlinks=true,
    citecolor=gray,
    filecolor=blue,
    linkcolor=blue,
    urlcolor=red
}
\usepackage{mathrsfs}

\definecolor{ICGMmarine}{rgb}{0.168,0.168,0.525}
\definecolor{ICGMblue}{rgb}{0.,0.549,0.714}
\definecolor{ICGMorange}{rgb}{0.968,0.647,0.}
\definecolor{ICGMyellow}{rgb}{1,0.804,0.}

\definecolor{dgreen}{rgb}{0,.5,0}
\definecolor{dred}{rgb}{.7,.0,.0}
\newcommand{\bmg}{{\bm{\gamma}}}
\newcommand{\bmn}{{\bm{\eta}}}
\newcommand{\bmnabla}{{\bm{\nabla}}}
\newcommand{\bfn}{{\bf{n}}}
\newcommand{\bfx}{{\bf{x}}}

\newcommand{\bfu}{{\bf{u}}}
\newcommand{\bfz}{{\bf{z}}}

\newcommand{\bfV}{{\bf{V}}}

\newcommand{\be}{\begin{eqnarray}}
\newcommand{\ee}{\end{eqnarray}}

\usepackage{xr}
\newcommand{\br}{\mathbf{r}}
\newcommand{\Hxc}{{\rm Hxc}}

\begin{document}

\title{Variational minimization scheme for the one-particle reduced density matrix  functional theory in the ensemble N-representability domain}
\author{Matthieu Vladaj}
\affiliation{ICGM, Universit\'{e} de Montpellier, CNRS, ENSCM, Montpellier, France}
\author{Quentin Mar\'ecat}
\affiliation{ICGM, Universit\'{e} de Montpellier, CNRS, ENSCM, Montpellier, France}
\author{Bruno Senjean}
\affiliation{ICGM, Universit\'{e} de Montpellier, CNRS, ENSCM, Montpellier, France}
\author{Matthieu Sauban\`ere}
\email{matthieu.saubanere@cnrs.fr}
\affiliation{ICGM, Universit\'{e} de Montpellier, CNRS, ENSCM, Montpellier, France}
\affiliation{Universit\'{e} de Bordeaux, CNRS, LOMA, UMR 5798, F-33400 Talence, France}

%%%%% Abstract %%%%%

\begin{abstract}
The one-particle reduced density-matrix (1-RDM) functional theory is a promising alternative to density-functional theory (DFT) that uses the
1-RDM rather than the electronic density as a basic variable.
However, long-standing challenges such as the lack of Kohn--Sham scheme and the complexity of the pure $N$-representability conditions are still impeding its wild utilization.
Fortunately, ensemble $N$-representability conditions derived in the natural orbital basis 
are known and trivial, such that almost every functionals of the 1-RDM are actually natural orbital functionals which do not perform
well for all the correlation regimes.
In this work, we propose a variational minimization scheme in
the ensemble $N$-representable domain
that is not restricted to the natural orbital representation of the 1-RDM.
We show that splitting the minimization into the diagonal and off-diagonal part of the 1-RDM can open the way toward the development of functionals of the orbital occupations,
which remains
a challenge for the generalization of site-occupation functional theory in chemistry.
Our approach is tested on the uniform Hubbard model using the M\"uller and the T\"ows--Pastor functionals, as well as on the dihydrogen molecule
using the M\"uller functional.
\end{abstract}

\maketitle

\section{Introduction}

Over the last two decades, density functional theory (DFT) has played a major role in the
study of the physico-chemical properties of molecules and materials.
The key idea behind DFT, formulated by Hohenberg and Kohn~\cite{hktheo}, is to replace the $N$-electron wavefunction by the electronic density as the fundamental variable of the many-electron problem.
Within the Kohn--Sham (KS) decomposition scheme~\cite{KS},
the electronic repulsion
effects are treated implicitly by a Hartree-exchange-correlation (Hxc) functional of the density, thus leading to a good ratio between
computational cost and accuracy.
%and extensions including system under magnetic field, finite-temperature or time-dependent systems [see Refs.~\cite{parr1989density,gross1995density} and references therein].
However, despite the numerous Hxc functionals developed
in the past decades ~\cite{marques2012libxc,mardirossian2017thirty},
KS-DFT still faces issues when dealing with systems that exhibit strong (or static) correlation effects ~\cite{cramer2009density,cohen2008insights}.
Improvements have been developed
recently by considering fractional spin and charge corrections~\cite{li2018localized,su2018describing},
or by using the many-electron expansion~\cite{zhu2016many}.
Another interesting approach is to construct the adiabatic connection from the interacting to the strong-interaction limit of KS-DFT~\cite{seidl1999strictly,seidl1999strong,gori2009density,
malet2012strong,friesecke2022strong}.

Alternatively,
replacing the density by the one-particle reduced density-matrix (1-RDM) as the fundamental
variable leads to the 1-RDM functional theory (1-RDMFT)~\cite{lowdin1955quantum,gilbert1975hohenberg,
coleman1963structure,haar1961theory}
and has several advantages.
First, kinetic and exchange energy contributions are expressed exactly and analytically with respect to the 1-RDM. 
Second, fractional occupations of the orbitals are naturally obtained within 1-RDMFT, thus allowing for the proper description of bond dissociation where static correlation effects are significant.
Several works have used the advantages
of 1- and 2-RDMFT to describe
strong correlation in the framework of KS-DFT~\cite{zhang2023progress,gibney2023universal}.
Despite these advantages, 1-RDMFT does not yet compete with DFT, partially due to the lack of an efficient scheme as provided by Kohn and Sham.
Indeed, the non-idempotent 1-RDM of an interacting system cannot be mapped to its non-interacting equivalent that is idempotent by construction, as it results from a single-electron Schr\"odinger equation~\cite{piris2017comment}.
Besides, the proper minimization of the energy functional of the 1-RDM is in principle highly non-trivial as it requires maintaining the {\it pure} $N$-representability conditions of the 1-RDM, also known as the 
generalized Pauli constraints (GPCs)~\cite{altunbulak2008pauli,
schilling2018generalized,castillo2023effective}).
Indeed, the number of GPCs increases significantly with the number of electrons and orbitals~\cite{theophilou2015generalized},
although it can be drastically reduced by introducing
physically motivated ansatz, as suggested by Gritsenko and Pernal~\cite{gritsenko2019approximating}.
In practice, the 1-RDM is usually written in the natural orbital (NO) representation,
and the minimization is performed via a Langrangian-based system of differential equations where both the NOs and their occupation numbers are optimized while satisfying the more trivial \textit{ensemble} $N$-representability conditions~\cite{coleman1963structure,
coleman1972necessary,valone1980consequences},
which can coincide with the
pure $N$-representable set of 1RDMs in some
specific cases~\cite{schilling2019diverging}.
Because these conditions are known in the NO basis,
Hxc functionals
are almost all NO-functionals
(NOFs)~\cite{muller1984explicit,csanyi2000tensor,csanyi2002improved,buijse2002approximate,
sharma2008reduced,lathiotakis2009functional, 
goedecker1998natural,holas1999properties,
cioslowski1999constraints,cioslowski20231}. 
The most recent and efficient ones are
the
Piris natural orbitals functionals (PNOFs)~\cite{piris2003one,piris2004natural,leiva2005natural,
piris2006new,piris2007dispersion,piris2009iterative,
piris2010accurate,piris2011natural,piris2013natural,
piris2014perspective,lopez2015performance}
based on the cumulants~\cite{kutzelnigg1999cumulant}.
Note that despite promising results, NOFs do not perform well for all the correlation regimes~\cite{ramos2015h4,cioslowski2015robust,
mitxelena2017performance,rodriguez2017comprehensive,
mitxelena2018phase}, although recent progress are worth mentioning~\cite{piris2021global}.

Solving the aforementioned non-linear differential equations is computationally demanding, so that alternative strategies have been developed to define efficient and robust algorithms to perform the energy functional minimization.
Gilbert proposed to determine the NOs using a non-local potential for functionals
that are explicitly dependent on the 1-RDM~\cite{gilbert1975hohenberg},
whose existence was later shown by Pernal in the case of NOFs~\cite{pernal2005effective}.
In this approach, partially occupied orbitals are all degenerate.
A faster scheme for the NO optimization is based on a hermitian matrix built from the Lagrange multiplier, as proposed by Piris and Ugalde~\cite{piris2009iterative}.
One can also use the Hessian and
couple the optimization of the NOs and occupation numbers~\cite{castillo2023effective}.
Another interesting scheme called the local 1-RDMFT was
developed by Lathiotakis and coworkers,
where the minimization of the NOs is restricted to a domain where the orbitals are eigenfunctions of a single-electron Hamiltonian with a local potential, thus
bridging the 1-RDMFT and KS-DFT approaches~\cite{lathiotakis2014local,theophilou2015orbitals}.
Written in the NO representation, these methods
as well as methods based on an analogy with the temperature
are particularly suited for NOFs~\cite{goedeckerEfficientLinearScaling1994,
baldsiefenMinimizationProcedureReduced2013,
saubanere2016interaction,
wangSelfConsistentFieldMethodCorrelated2022}.
Indeed, in that case the energy does only depend explicitly on the occupation numbers while
the dependence on the NOs is implicitly accounted to in the Hartree and exchange integrals.
Alternatively,
minimization schemes based on a different representation than natural orbitals were also proposed, for instance 
using projected gradient-based algorithms constrained to preserve ensemble $N$-representability conditions~\cite{cancesProjectedGradientAlgorithms2008}.

In this work, we introduce a new variational minimization scheme so-called density-matrix interpolation variational ansatz (DIVA) which relies on remarkable topological properties of the ensemble $N$-representable set of 1-RDM.
In contrast to the usual minimization schemes adapted to
NOFs, DIVA aims to minimize the energy functional of the 1-RDM expressed in any orthonormal orbital basis, thus
motivating the development of new 1-RDM functionals beyond
NOFs.
By separating the minimization over
the diagonal and off-diagonal part of the 1-RDM,
one can also make a connection between 
1-RDMFT and DFT in a lattice, also called site-occupation functional theory (SOFT)~\cite{gunnarsson1986density,DFT_lattice,
lima2002density,lima2003density}.
We show how this separation can be used to estimate Hartree Hxc potentials,
thus paving the way toward the development of functionals of 
the orbital occupations which is the current limitation for the implementation of SOFT in chemistry~\cite{fromager2015exact,coe2019lattice}.

The paper is organized as follows.
In Sec.~\ref{sec:RDMFT}, we give a brief introduction to 1-RDMFT. The separation of the minimization process into
that bridges 1-RDMFT and SOFT
is then presented in Sec.~\ref{sec:SOFT-DIVA}.
Our variational minimization strategy and its various implementations are described in Sec.~\ref{sec:DIVA},
and tested on the Hubbard model and the hydrogen molecule
in Sec.~\ref{sec:results}.
Finally, conclusions and perspectives are provided in Sec.~\ref{sec:conclu}.
Note that in the following, we will use `1-RDM' and `density matrix' to refer to the same object, i.e. the one-particle reduced density matrix.

\section{Reduced density-matrix functional theory}
\label{sec:RDMFT}

Let us consider the second-quantized electronic Hamiltonian written in an orthonormal basis set, referred to as the computational basis set and composed of $N$ spin-orbitals $\lbrace \varphi_i(\bfx) \rbrace$ that are different from the set of natural spin-orbitals $\lbrace \phi_i(\bfx) \rbrace$,
\begin{eqnarray}\label{eq:H_elec}
\hat{H} &=&
\hat{T} + \hat{U} \nonumber \\
%&=&  \sum_{ij}^{N} \sum_{\sigma} t_{ij\sigma} \hat{c}_{i\sigma}^{\dagger}\hat{c}_{j\sigma} + \sum_{ijkl}^{N}\sum_{\sigma \sigma'} U_{ijkl\sigma \sigma'}\hat{c}_{i\sigma}^{\dagger}\hat{c}_{k\sigma}\hat{c}_{j\sigma'}^{\dagger}\hat{c}_{l\sigma'},
&=&  \sum_{ij}^{N} t_{ij} \hat{c}_{i}^{\dagger}\hat{c}_{j} + \dfrac{1}{2}\sum_{ijkl}^{N}U_{ijkl}\hat{c}_{i}^{\dagger}\hat{c}_{j}^\dagger\hat{c}_{k}\hat{c}_{l},
\end{eqnarray}
%where $\hat{c}_{i\sigma}^{\dagger}$ ($\hat{c}_{i\sigma}$) corresponds to the creation (annihilation) of an electron with spin $\sigma$ on orbital $i$. The one-body integrals $t_{ij\sigma}$ contain the kinetic and external potential contributions, while the two-body integrals $U_{ijkl\sigma\sigma'}$ contain contributions from the electron-electron interaction.
where $\hat{c}_{i}^{\dagger}$ ($\hat{c}_{i}$) corresponds to the creation (annihilation) of an electron in $\varphi_i(\bfx)$. 
The one-body integrals
\begin{eqnarray}\label{eq:one-body}
t_{ij} = \int {\rm d}\bfx \varphi_i^*(\bfx)\left(
- \dfrac{\nabla^2}{2} + \sum_I \dfrac{Z_I}{| \br - {\bf R}_I |} \right) \varphi_j(\bfx)
\end{eqnarray}
contain the kinetic and external potential contributions, while the two-body integrals
\begin{eqnarray}
U_{ijkl} = \int {\rm d}\bfx_1 {\rm d}\bfx_2
\dfrac{ \varphi_i^*(\bfx_1) \varphi_j^*(\bfx_2) \varphi_k(\bfx_2) \varphi_l(\bfx_1)}{|\br_1 - \br_2 |}
\end{eqnarray}
contain contributions from the electron-electron interactions. 
The summation in Eq.~(\ref{eq:one-body})
runs over the number of atoms at positions ${\mathbf{R}_I}$ and $\bfx_i = (\br_i,\sigma_i)$ corresponds to the position $\br_i$ and spin $\sigma_i$ of the electron $i$.

In 1-RDMFT, the basic variable $\bmg$ is the 1-RDM given by
\begin{eqnarray}
%  \gamma_{ij\sigma} = \langle \Psi | \hat{c}_{i \sigma}^\dagger \hat{c}_{j \sigma} | \Psi \rangle ,
\gamma_{ij} = \langle \Psi | \hat{c}_{i}^\dagger \hat{c}_{j} | \Psi \rangle ,
\end{eqnarray}
with $|\Psi\rangle$ the ground-state wavefunction of $\hat{H}$.
The energy can in principle be written as a functional of the 1-RDM,
\begin{eqnarray}\label{eq:1st_decomp}
E[\bmg] = T[\bmg] + W[\bmg],
\end{eqnarray}
where
\begin{eqnarray}
%T[\bmg] = \sum_{ij,\sigma} t_{ij\sigma} \gamma_{ij\sigma}
T[\bmg] = \sum_{ij} t_{ij} \gamma_{ij}
\end{eqnarray}
is an explicit
functional of the 1-RDM
and 
$
W[\bmg]
$
is the energy functional describing the electronic repulsion.
The analytical form of this functional remains unknown and is usually approximated, though it
can be written using the Levy--Lieb formalism as follows,
\begin{eqnarray}
W[\bmg] = \min_{\Psi \rightarrow \bmg} \bra{\Psi} \hat{U} \ket{\Psi},
\end{eqnarray}
where the notation $\Psi \rightarrow \bmg$ refers to any wavefunction giving the one-particle density matrix $\bmg$.
Within the variational principle of 1-RDMFT, the ground-state energy $E_{0}$ can be obtained as a saddle point,
\begin{eqnarray}
  E_{\rm 0} = \min_{\bmg} E[\bmg].
  \label{eq:saddle_point}
\end{eqnarray}
More precisely, 
as a corollary of the Hohenberg--Kohn theorem (later extended by Gilbert~\cite{gilbert1975hohenberg} to non-local potential), 
there exists a one-to-one mapping between the ground-state wavefunction and the 1-RDM, thus allowing to determine the ground-state energy variationally.
The conditions for $\bmg$ to derive from a $N_e$-electron wavefunction -- so-called $N$-representability conditions -- are highly non-trivial~\cite{altunbulak2008pauli,
schilling2018generalized,castillo2023effective} and cannot be used in practice to define the domain in which the minimization of Eq.~(\ref{eq:saddle_point}) takes place.
However, the conditions for $\bmg$ to be ensemble $N$-representable are known and straightforward~\cite{coleman1963structure}
and are given as follows, 
\begin{equation}
\label{Coleman_const}
\begin{split}
        \sum_{k=1}^N \eta_k = N_e, \;\;
        0 \leq \eta_k \leq 1, 
\end{split}
\end{equation}
where $\eta_k$ is the occupation number of the $k$-th natural orbital.
Hence, it is common to use
the natural orbital representation of the density
matrix, i.e. $\bmg \equiv (\lbrace \eta_k \rbrace, \lbrace \ket{\phi_k}\rbrace)$.
As originally proposed by Gilbert~\cite{gilbert1975hohenberg}, these conditions are fulfilled during the minimization by introducing a Lagrangian function in the ensemble $N$-representable domain,
\begin{eqnarray}
\label{Lagrangian}
     \mathcal{E} & = & E[\lbrace \eta_k \rbrace, \lbrace \ket{\phi_k}\rbrace] - \mu \bigg(\sum_k \cos^2(\theta_k) - N_e \bigg)  \nonumber \\
  && - \sum_{jk} \lambda_{jk} \bigg( \langle \phi_j | \phi_k \rangle - \delta_{jk} \bigg),
\end{eqnarray}
where $\mu$ and $\lbrace \lambda_{jk}\rbrace$ are Lagrange multipliers, the first term ($\eta_k = \cos^2(\theta_k)$) fulfills Eq.~\eqref{Coleman_const} and the second term the orthonormality of the natural orbitals $\lbrace | \phi_k \rangle \rbrace$.
Minimizing Eq.~\eqref{Lagrangian} with respect to the whole set of occupation numbers and NOs leads to two non-linear differential equations, 
\begin{eqnarray}
    \dfrac{\partial \mathcal{E}}{\partial \theta_k} = \left(\mu-\dfrac{\partial E}{\partial \eta_k} \right) \sin(2 \theta_k) = 0,
\end{eqnarray}
and
\begin{eqnarray}
 \displaystyle     \dfrac{\partial \mathcal{E}}{\partial \langle \phi_k|} = \dfrac{\partial E}{\partial \langle \phi_k| } - \sum_j \lambda_{jk} | \phi_j \rangle = 0,
\end{eqnarray}
that are solved successively but with slow convergence~\cite{cartier2024exploiting}.
This situation contrasts with the variational equations
in KS-DFT and SOFT that are solved efficiently through the definition of a single-particle Schr\"odinger equation~\cite{KS,gunnarsson1986density}.

\section{Link between 1-RDMFT and SOFT}
\label{sec:SOFT-DIVA}

Rather than decomposing the 1-RDM into
natural orbitals and occupation numbers,
we decompose it in terms of its diagonal elements,
i.e. orbital (or site) occupations,
$\bfn \equiv \lbrace \gamma_{ii} \rbrace$
 and the $N(N-1)$ (not nil in general) off-diagonal elements denoted by $\bmg^{\neq}$ such that $\bmg = \bfn \cdot \mathbb{1} + \bmg^{\neq}$,  where $\mathbb{1}$ refers to the identity matrix. 
Note that $Tr.(\bfn) = Tr.(\boldsymbol{\eta}) = N_e$, the number of electrons, and $n_i = \sum_k \eta_k |u_{ik}|^2$ where $\bfu$ corresponds to the eigenvector matrix of $\bmg$.
Hence, $\bfn$ is a vector of size $N$ that by definition differs from the occupation number vector $\boldsymbol{\eta}$, for instance each $n_i$ can be fractional even if $\bmg$ is idempotent.
The energy functional in
Eq.~(\ref{eq:1st_decomp}) therefore
reads
  \begin{equation}
    E[\bmg]  =  T[\bfn,\bmg^{\neq}] + W[\bfn,\bmg^{\neq}]
    \label{eq:Edeg}.
  \end{equation}
Following the KS decomposition, Eq.~(\ref{eq:Edeg}) can be alternatively written as
\begin{eqnarray}
  E[\bmg] =  T_s[\bfn] +  F[\bfn,\bmg^{\neq}],
\end{eqnarray}
where $T_s[\bfn]$ represents the one-body (kinetic and potential) energy contributions of the non-interacting system with orbital occupations $\bfn$, i.e. $T_s[\bfn] = \langle \Phi | \hat{T} | \Phi \rangle $ differs from $T[\bmg]$ as $|\Phi \rangle$ corresponds to a single Slater determinant such that $\langle \Phi | \hat{\bfn} | \Phi \rangle  = \bfn$.
$F[\bfn,\bmg^{\neq}]$ is a functional containing all non trivial terms defined by
\begin{eqnarray}
  \label{eq:def_F}
  F[\bfn,\bmg^\neq]   =  T[\bfn,\bmg^\neq] - T_s[\bfn]+ W[\bfn,\bmg^\neq].
  \end{eqnarray}
In analogy with the natural orbital representation, the minimization of the energy leads to two saddle-point equations.
The first saddle-point equation reads
\begin{eqnarray}
  \delta \bfn \left\lbrace \frac{\delta T_s[\bfn]}{\delta \bfn}  + \left.\frac{\delta F[\bfn, \bmg^\neq]}{\delta \bfn}\right|_{\bmg^\neq} \right\rbrace = 0,
\end{eqnarray}
and leads to a pseudo Kohn-Sham (pKS) equation:
\begin{eqnarray}
  \hat{h}_{\rm pKS}[\bfn] |\Phi^{\rm pKS} \rangle =  \left(   \hat{T} + \hat{V}_{\Hxc}[\bfn] \right)|\Phi^{\rm pKS}\rangle = \mathcal{E}^{\rm pKS} |\Phi^{\rm pKS}\rangle.
    \label{eq:KS_type1}
\end{eqnarray}
Eq.~(\ref{eq:KS_type1}) is similar to the self-consistent equation in SOFT, where $\hat{h}_{\rm pKS} [\bfn]$  corresponds to an effective non-interacting Hamiltonian. 
However, in contrast to SOFT, 
the Hxc potential $\bfV_{\Hxc}[\bfn]$ is defined through the full interacting 1-RDM, at the saddle point of $\bmg^{\neq}$,
\begin{eqnarray}
  \bfV_{\Hxc}[\bfn] := \left.\bfV_{\rm Hxc}[\bfn]\right|_{\bmg^{\neq}} = \left. \frac{\delta F[\bfn, \bmg^{\neq}]}{\delta \bfn}\right|_{\bmg^{\neq}}.
   \label{eq:KS_VHXC}
\end{eqnarray}
$\bfV_{\Hxc}[\bfn]$  is an on-site (diagonal) one-body potential that imposes
the equality between the effective chemical potential $\mu^{\rm pKS}$ and the interacting chemical potential $\mu = \partial E/\partial \bfn$
at convergence, or, in other words, $n_i^{\rm pKS} =  \gamma_{ii}^{\rm pKS} = \langle \Phi | \hat{c}_i^\dagger \hat{c}_i | \Phi \rangle = n_i$.
Note however that the off-diagonal elements of the idempotent 1RDM associated to the pKS Hamiltonian differs from the off-diagonal element of the interacting non-idempotent 1-RDM, $\gamma_{ij}^{\rm pKS} =  \langle \Phi^{\rm pKS} | \hat{c}_{i}^\dagger\hat{c}_{j} | \Phi^{\rm pKS} \rangle \neq \gamma_{ij}$ if $i\neq j$.
% In other-words, the non-interacting pKS system, associated with an idempotente 1RDM, is self-consistently defined, through $\bfV_{\Hxc}[\bfn]$, in order to share (fractional) OAO/site occupations than the interacting system associated with an non-idempotent 1-RDM.
As in SOFT, only the chemical potential and site (or orbital) occupations of the interacting system should be recovered from the pKS one.
Therefore, 1RDMs and more specifically natural orbitals and occupation numbers are different in both systems.
Eq.~(\ref{eq:KS_type1}) is determined concomitantly with the complementary saddle point equation:
\begin{eqnarray}
  \label{eq:saddle_gd}
  \delta \bmg^{\neq} \left\lbrace \left.\frac{\delta F[\bfn, \bmg^{\neq}]}{\delta \bmg^{\neq}}\right|_{\bfn = \bfn^{\rm pKS}} \right\rbrace= 0,
\end{eqnarray}
where $\bfn^{\rm pKS}$ are the orbital/site occupations resulting from Eq.~(\ref{eq:KS_type1}).
In practice, solving Eq.~(\ref{eq:KS_type1}) appears numerically easy when $\bfV_{\Hxc}[\bfn]$
is known,
however solving Eq.~(\ref{eq:saddle_gd}) remains difficult, in particular due to the non-trivial representability conditions that
$\bmg^{\neq}$ has to fulfill. In Sec.~\ref{sec:SOFT_DIVA}, we propose an algorithm to determine $\bfV_{\Hxc}[\bfn]$ from  Eq.~(\ref{eq:saddle_gd}) such that Eqs.~(\ref{eq:KS_type1}) and~(\ref{eq:saddle_gd}) can be solved iteratively.

From a more conceptual perspective, one can also use a Levy--Lieb formalism to derive similar equations,
\begin{eqnarray}
  E_{\rm 0} &= &\min_{\bfn}\left\{ \min_{\bmg \rightarrow \bfn} \left\{ E[\bfn, \bmg^{\neq}] \right\}\right\} \nonumber \\
& =& \min_{\bfn}\left\{ \min_{\bmg \rightarrow \bfn} \left\{T_s[\bfn]  + F[\bfn, \bmg^{\neq}] \right\}\right\}\nonumber \\
& =& \min_{\bfn}\left\{ T_s[\bfn]  +  \min_{\bmg \rightarrow \bfn} \left\{ F[\bfn, \bmg^{\neq}] \right\}\right\},
\end{eqnarray}
where $\bmg \rightarrow \bfn$ corresponds to density matrices with orbital occupations $\bfn$.
In that framework, the Hxc potential is defined as
\begin{eqnarray}
  \bfV_{\Hxc}[\bfn] =\frac{\delta}{\delta \bfn} \min_{\bmg \rightarrow \bfn} \left\{  F[\bfn, \bmg^{\neq}] \right\},
  \label{eq:extract_KS}
\end{eqnarray}
thus bridging 1-RDMFT and SOFT, where
the Hxc potential functional of the orbital occupations can be obtained after solving Eq.~(\ref{eq:saddle_gd}) for a fixed orbital occupation $\bfn$.

\section{Density-matrix Interpolation Variational {\it ansatz}}
\label{sec:DIVA}

\subsection{Topological properties of the  ensemble $N$-representable domain of the 1-RDM}

Let us consider the space $\mathcal{M}_N(\mathbb{R})$ of $N\times N$ symmetric and real matrices, 
$N$ being the number of spin-orbitals in the system. 
In this section we denote by $\Omega_\bmg$ the domain of  ensemble $N$-representable density matrices 
that is included in $\mathcal{M}_N(\mathbb{R})$, $\Omega_\bmg \subset \mathcal{M}_N(\mathbb{R})$.  
A significant advantage of working with natural orbitals
is that the 
 ensemble $N$-representable conditions are straightforward, i.e. a matrix $\bmg \in \mathcal{M}_N(\mathbb{R}) $ is ensemble $N$-representable if all the occupation numbers
$\eta_k$ fulfill 
$0 \leq \eta_{k} \leq 1$.
Similarly, it means that $\bmg$ is ensemble $N$-representable if both $\bmg$ and $\mathbb{1} -\bmg$ are positive semi-definite, i.e. both showing positive or null eigenvalues $\eta_k  \geq 0$ and  $1 - \eta_k \geq 0$ for all $1\leq k \leq N$. 
Hence, the domain of   ensemble $N$-representable density matrices reads
\begin{align}\label{eq:positive_semidef}
\Omega_{\bmg} = \left\lbrace \bmg \in \mathcal{M}_N(\mathbb{R}) \; | \; \eta_k \geq 0 \; \& \; 1-\eta_k \geq 0,\;  \forall k \right\rbrace.
 \end{align}
Interesting topological properties for the  ensemble $N$-representable 1-RDM domain $\Omega_{\bmg}$ follow from these definitions.
The first remarkable property of $\Omega_{\bmg}$  is the {\it convexity}, see proofs in Appendix~\ref{App:demo-convex}.
It follows that given a set of  ensemble $N$-representable matrices $\left\{\bmg^{(s)} \in \Omega_{\bmg}\right\}$, all matrices constructed as 
\begin{equation}
\label{eq:convexity}
\bmg(\bfz) = \sum_{s} z_s \bmg^{(s)}
\end{equation}
also belong to $\Omega_{\bmg}$ if all $z_s \geq 0$ and $\sum_s z_s = 1$.

%Interior
Let us now define $\mathring{\Omega}_{\bmg}$ the ensemble of matrices $\bmg$ such that both $\bmg$ and $\mathbb{1} -\bmg$ are positive definite (in contrast to Eq.~(\ref{eq:positive_semidef}) where they are positive semi-definite), i.e.
\begin{align}\label{eq:positive_def}
\mathring{\Omega}_{\bmg} = \left\lbrace \bmg \in \mathcal{M}_N(\mathbb{R})\; | \; \eta_k > 0 \; \& \; 1 - \eta_k > 0,\; \forall k \right\rbrace.
 \end{align}
 The convexity of $\mathring{\Omega}_{\bmg}$ follows the same proof as in Appendix~\ref{App:demo-convex} for $\Omega_{\bmg}$.
% Frontier
Besides, we introduce
 \begin{align}\label{eq:boundary}
  \delta \Omega_{\bmg} = \Omega_{\bmg} \, \backslash \, \mathring{\Omega}_{\bmg},
  \end{align}
such that $\delta \Omega_{\bmg} $ contains the density matrices $\bmg$ that exhibit at least one integer eigenvalue $\eta_k = \{0,1\}$. 
Contrary to $\Omega_{\bmg}$ and $\mathring{\Omega}_{\bmg}$,  $\delta \Omega_{\bmg}$ is not convex.
We show in Appendix~\ref{App:demo-topo} that $\mathring{\Omega}_{\bmg}$ corresponds to the topological interior of  $\Omega_{\bmg}$, $\delta \Omega_{\bmg}$ is the topological boundary of $\Omega_{\bmg}$ and $\Omega_{\bmg} = \mathring{\Omega}_{\bmg} \cup \delta \Omega_{\bmg}$ is a topological closed set. 
\\

  % space Idempotent
Finally, let us define $\delta \Omega_{\bmg}^* \subset \delta \Omega_{\bmg} $ the set of idempotent matrices, i.e. $\bmg = {\bmg}^2$ if $\bmg \in \delta\Omega_{\bmg}^*$
and
 \begin{align}
  \delta \Omega_{\bmg}^* = \left\lbrace \bmg \in \Omega_{\bmg} \,| \,  \eta_k = \left \{ 0, 1\right\},\, \forall k \right\rbrace, 
 \end{align}
which is also non convex.
$\delta \Omega_{\bmg}^*$ corresponds to the set of extreme points of $\Omega_{\bmg}$, i.e. given $\bmg_* \in \delta\Omega_{\bmg}^*$ there is no $\bmg^A$ and $\bmg^B$ in $\Omega_{\bmg}$ such that $\bmg_* = z \bmg^A +(1-z)\bmg^B$ with $0 <z<1$. Following the Krein--Milman theorem,
there exists a decomposition in terms of idempotent matrices, i.e. for any $\bmg \in \Omega_\bmg$,
 \begin{equation}
   \bmg(\bfz) = \sum_s z_s \bmg^{(s)}_{*},
   \label{Eq:DIVA_1}
 \end{equation}
 with  $\bfz = \lbrace z_s \rbrace$, $\sum_s z_s = 1$ and  $\bmg^{(s)}_* \in \delta \Omega_{\bmg}^*$.
Given Eqs.~(\ref{eq:saddle_point}) and (\ref{Eq:DIVA_1}),
it follows the variational ansatz:
\begin{equation}
  E_{\rm 0} = \min_{\bfz} E[\bmg(\bfz)].
  \label{Eq:DIVA_gs}
\end{equation}
Hence, the energy is minimized with respect to the weights $\bfz$ defining the interpolation of $\bmg(\bfz)$
in terms of idempotent 1-RDM.
We show in Appendix~\ref{App:demo-decomp} that for each $\bmg \in \mathring{\Omega}_{\bmg}$, the sum in Eq.~(\ref{Eq:DIVA_1}) runs up to $2^N$ idempotent matrices. It follows that the number of variational parameters rapidly becomes prohibitive when the size of the system increases. 
To reduce the number of parameters, one could define the trial set as every Slater determinants corresponding to the singly- and doubly-excited determinants with respect to the Hartree--Fock state. 
The efficiency and accuracy of such ansatz strongly depends on the size and relevance of the trial set of $\lbrace \bmg_{*}^{(s)} \rbrace$.

Rather than interpolating $\bmg$ in terms of
idempotent 1-RDM, we can use the property that for each $\bmg \in \Omega_{\bmg}$ there exists a decomposition in terms of {\it boundary} matrices
belonging to $\delta \Omega_\bmg$ [see Eq.~(\ref{eq:boundary})].
Then, Eq.~(\ref{Eq:DIVA_1}) turns to
\begin{equation}
  \bmg(\bfz) = \sum_s z_s \bmg_B^{(s)}.
  \label{Eq:DIVA_3}
\end{equation}
As $\bmg_B^{(s)}$ can be seen as a linear combination of $\bmg_*^{(s)}$, one can expect the amount of variational parameters $\lbrace z_s \rbrace$ to be drastically reduced
when too much $\lbrace \bmg_*^{(s)} \rbrace$ would be needed to reproduce $\bmg$. 
Indeed, any non-idempotent $\bmg \in \mathring{\Omega}_\bmg$ can in principle be obtained
by interpolating only two boundary matrices.
Hence,
while the set of idempotent matrices $\left\{\bmg^{(s)}_*\right\}$ in Eq.~(\ref{Eq:DIVA_1}) is
very large but straightforward to construct, 
only a small relevant set of boundary matrices  $\left\{\bmg_B^{(s)}\right\}$ would suffice but is non-trivial to find.
An efficient algorithm would consist in constructing this non-trivial set of relevant $\left\{\bmg_B^{(s)}\right\}$ on which the minimization in Eq.~(\ref{Eq:DIVA_gs}) is performed.
Eqs.~(\ref{Eq:DIVA_gs}) and (\ref{Eq:DIVA_3}) constitute the density-matrix interpolation variational ansatz (DIVA). 
Note that by construction, all density matrices of the form given by Eq.~(\ref{Eq:DIVA_3}) are ensemble $N$-representable.

Interestingly, the topological properties presented in $\mathcal{M}_N(\mathbb{R})$  also hold if we restrain the considered metric space to  $\mathcal{M}^{N_e}_N(\mathbb{R})$ and $\mathcal{M}^{\bfn}_N(\mathbb{R})$, corresponding to the spaces of 
$N \times N$ symmetric and real matrices with a fixed trace (fixed number of electron $N_e$) and fixed diagonal part $\bfn$, respectively.
Considering these restrictions should automatically reduce the number of trial 1-RDMs in Eq.~(\ref{Eq:DIVA_3}).

\subsection{Iterative gradient-based implementations of DIVA}\label{sec:implementations}

%In the previous section, it has been shown that the set $\Omega_{\gamma}$ is convex and closed. Taking advantage of these properties, on can express any $\gamma \in \Omega_{\gamma}$ in the form:
%
%\begin{equation} \label{eq convexity}
%    \gamma = \sum_i z_i \gamma_i
%\end{equation}
%where $\gamma_i \in \Omega_{\gamma} \forall i$ and $z_i \in [0,1]$ 

\begin{figure}
    \centering
      \resizebox{\columnwidth}{!}{
      \includegraphics[scale=1]{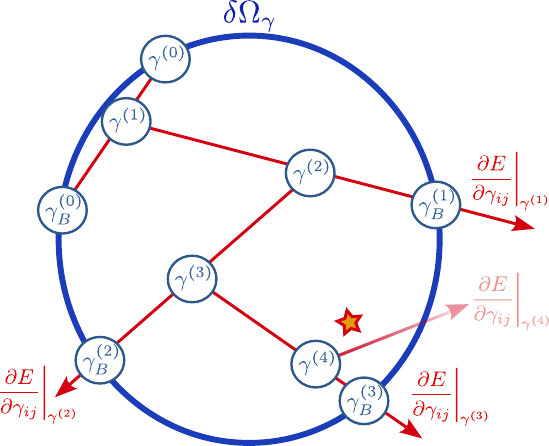}
        }
    \caption{Illustration of the mono-parameter DIVA. The set $\Omega_{\bmg}$ is illustrated by the union of $\delta \Omega_{\bmg}$ (blue circle) and the topological interior. Red arrows represent the direction of the energy gradient, used to find the $\bmg_B^{(s)}$ belonging to the border of the representability domain ($\delta\Omega_\bmg$). Each $\bmg^{(s)}$ is obtained as an interpolation between $\bmg^{(s-1)}$ and $\bmg_B^{(s-1)}$. 
The star represents the ground-state 1-RDM $\bmg_0$ (at the level of the functional approximation)
that is recovered after convergence.}
    \label{fig:monoparam}
\end{figure}

\begin{figure}
    \centering
      \resizebox{\columnwidth}{!}{
      \includegraphics[scale=1]{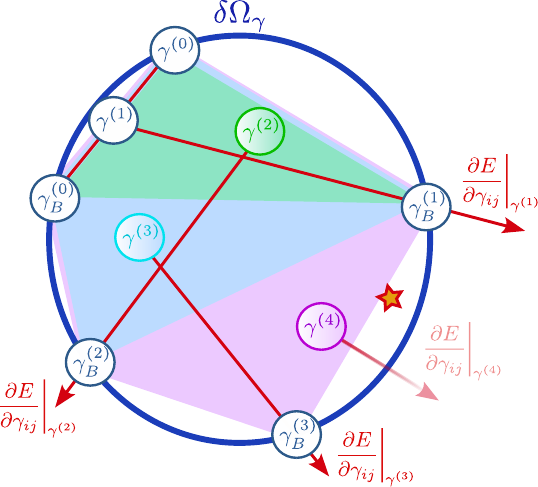}
        }
    \caption{Illustration of the multi-parameter DIVA. The set $\Omega_{\bmg}$ is illustrated by the union of $\delta \Omega_{\bmg}$ (blue circle) and the topological interior. Red arrows represent the direction of the energy gradient, used to find the $\bmg_B^{(s)}$ belonging to the border of the representability domain ($\delta\Omega_\bmg$). Each $\bmg^{(s>2)}$ is a linear combination of all $\bmg_B^{(r)}$ with $r<s$. The star represents the ground-state 1-RDM $\bmg_0$ (at the level of the functional approximation) that is recovered after convergence.}
    \label{fig:multiparam}
\end{figure}

Let us now turn to the numerical implementations of DIVA.
Starting from a guess $\bmg^{(0)}$ that already belongs to $\Omega_{\bmg}$, 
we want to construct iteratively a minimal trial set of boundary density matrices to minimize the 1-RDMFT energy [Eq.~(\ref{Eq:DIVA_gs})]
according to Eq.~(\ref{Eq:DIVA_3}),
\begin{equation}
    \bmg(\bfz) = z_0 \bmg^{(0)} + \sum_{s} z_{s} \bmg_{B}^{(s)}.
\end{equation}
In the following, we highlight two ways of minimizing the energy functional within DIVA.
The first one consists in minimizing the energy functional
with respect to only one variational parameter $z_i$ at each iteration $i$ and is called the mono-parameter DIVA.
For the second one, the minimization is done over the whole set of parameters $\lbrace z_r \rbrace_{r\leq s}$
at iteration $s$, and is called the multi-parameter DIVA. These approaches are pictured in Figs.~\ref{fig:monoparam} and \ref{fig:multiparam} and detailed below.

\subsubsection{Mono-parameter DIVA}\label{sec:monoDIVA}

The mono-parameter DIVA is implemented according to the following steps.
\begin{enumerate}
\item Initialization (iteration $s=0$): set $\bmg^{(0)}$ to be the
(idempotent) density matrix coming from an Hartree--Fock or KS-DFT calculation,
thus ensuring that $\bmg^{(0)}$ belongs to $\Omega_{\bmg}$,
and even to $\delta \Omega_\bmg^* \subset \delta \Omega_\bmg \subset \Omega_\bmg$. 
To avoid divergence of the energy gradient
with respect to $\bmg^{(0)}$ having occupation numbers equal to 0, which happens for many NOFs, the first gradient is set to $\nabla_{ij}[\bmg] = (1- \delta_{ij})  \gamma_{ij}^{(0)}$.

\item Search the boundary density matrix $\bmg_B^{(s)} \in \delta\Omega_{\bmg}$
following the opposite direction of the energy gradient. This is done in two steps:
  \begin{enumerate}
  \item search for $\bmg_{\rm out} = \bmg^{(s)} - \theta \bmnabla[\bmg^{(s)}]$ by incrementing $\theta$ until $\bmg_{\rm out} \notin \Omega_{\bmg}$
  \item Once $\bmg_{\rm out}$ is found, vary $\theta$ using a bracketing method until $\bmg_{\rm out} \in \delta \Omega_{\bmg}$ and set $\bmg_B^{(s)} = \bmg_{\rm out}$
  \end{enumerate}
\item start a new iteration, $s = s + 1$
\item Interpolate between $\bmg^{(s-1)}$ and $\bmg_B^{(s-1)}$ to find $\bmg^{(s)} = (1-z_s)\bmg^{(s-1)} + z_s \bmg_B^{(s-1)}$ such that
  \begin{equation}
   \dfrac{\partial E[\bmg^{(s)}]}{\partial z_s} = 0.
  \end{equation}
\item Compute the energy gradient $\bmnabla[\bmg^{(s)}]$. Note that
the number of electrons is conserved by setting the diagonal part $\nabla_{ii}[\bmg^{(s)}] = \nabla_{ii}[\bmg^{(s)}] -\mu^{(s)}$ with $\mu^{(s)} = (1/N)\sum_i \nabla_{ii}[\bmg^{(s)}] $.
If the orbital occupations should be kept unchanged, set $\nabla_{ii}[\bmg^{(s)}] = 0$, $\forall i$.
Note that at convergence, $\mu^{(s)}$ corresponds to the chemical potential and $\nabla_{ii}[\bmg^{(s)}] -\mu^{(s)} \rightarrow 0$,
thus reflecting the equalization of the chemical potentials in the system.
\item If  $\delta E = |E^{(s-1)} - E^{(s)} | < \upsilon$ and $\delta \bmg = |\bmg^{(s-1)} - \bmg^{(s)} | < \varepsilon$
where $\upsilon$ and $\varepsilon$ are the convergence criteria, set $\bmg^{\rm opt} = \bmg^{(s)}$, which should be equal
to the ground-state 1-RDM $\bmg_0$ corresponding
to the level of approximation of the chosen 1-RDM functional (the star in Fig.~\ref{fig:monoparam}). 
Else, go to step 2.
\end{enumerate}

At convergence, the density matrix fulfills Eq.~(\ref{Eq:DIVA_3}), $\bmg(\bfz) = \sum_s^{M} Z_s \bmg_B^{(s)}$, with $Z_s = z_{s+1}\prod_{s+2}^{M}(1-z_s)$ and $M$ the number of iterations needed to achieve convergence.
Following Ref.~\cite{cancesProjectedGradientAlgorithms2008},
step 2 of the algorithm
could be made more efficient by
projecting $\bmg^{(s)}$ to the boundary domain $\delta \Omega_\bmg$.
However, it would lead to a $\bmg_{B}^{(s)}$ that deviates from the direction pointed by the gradient.
Also, the opposite energy gradient direction can be replaced by the conjugate gradient.
This is what we implemented in this work,
as we observed a much faster convergence using it.

\subsubsection{Multi-parameter DIVA}\label{sec:multiDIVA}

Apart from step 4, the multi-parameter approach remains the same as the mono-parameter one.
Rather than doing the interpolation between $\bmg^{(s-1)}$ and $\bmg_B^{(s-1)}$ only,
the interpolation is done over the whole set
of previously determined $\bmg_B^{(r)}$ with $r<s$:
\begin{equation}
    \bmg^{(s)} = z^{(0)} \bmg^{(0)} + \sum_{r=1}^{s-1} z_r\bmg_B^{(r)}.
\end{equation}
The minimization is done over the whole set of parameters $\bfz = (z_0, \dots, z_{(s-1)})$,
\begin{equation}
    \dfrac{\partial E[\bmg^{(s)}]}{\partial \bfz} = 0.
\end{equation}
In contrast to the previous implementation, we expect to achieve a much faster convergence at the expense of more expensive optimizations of the $\bfz$ parameters.

\subsubsection{SOFT-DIVA}
\label{sec:SOFT_DIVA}

According to Sec.~\ref{sec:SOFT-DIVA} the variational minimization of the energy can be split in two iterative sub-processes.
The first sub-process optimizes orbital/site occupations $\bfn$ through a pseudo-KS eigenvalue problem [see Eq.~(\ref{eq:KS_type1})], while the second sub-process 
optimizes the off-diagonal elements of the 1-RDM $\bmg^{\neq}$ [see Eq.~(\ref{eq:saddle_gd})]
and can be performed within the DIVA framework. 
Consequently the SOFT-DIVA algorithm can be implemented by performing the following steps.
\begin{enumerate}
\item Initialize $\bfV_{\Hxc}[\bfn]^{(s)}$ at iteration $s=1$. In practice we set the diagonal potential $\bfV_{\Hxc}[\bfn]^{(s)}$ to be equal to the diagonal part of the XC potential obtained from standard DFT approximation.
\item Plug $\bfV_{\Hxc}[\bfn]^{(s)}$ into the single-particle Hamiltonian (or Fock matrix) to generate a new idempotent 1-RDM $\bmg^{{\rm pKS}(s)}$ and thus a new set of orbital occupations $\bfn^{{\rm pKS}(s)} \equiv \lbrace \gamma_{ii}^{{\rm pKS}(s)}\rbrace$. The associated chemical potential $\mu^{{\rm pKS}(s)}$ can be defined as the highest occupied molecular orbital energy of the pKS system.
\item Optimize off diagonal element $\bmg^{\neq (s)}$, see Eq.~(\ref{eq:saddle_gd}). To that aim, initialize with $\boldsymbol{\gamma}^{\rm pKS}$ a DIVA process as described in Sec.~\ref{sec:monoDIVA} or \ref{sec:multiDIVA}
and set the diagonal part of the energy gradient to zero, such that the orbital occupations remain fixed and equal to 
$\bfn^{{\rm pKS}(s)}$ (see step 5 in Sec.~\ref{sec:monoDIVA})
along the full DIVA process.
DIVA will converge to an optimal density matrix of the form $\bmg^{(s)} = \bfn^{\rm pKS(s)} \cdot \mathbb{1} + \bmg^{\neq (s)}$.
\item Self-consistent condition. Compute the energy gradient vector ${\rm diag}(\bmnabla[\bmg^{(s)}]) = \lbrace \nabla_{ii}[\bmg^{(s)}] \rbrace_{i} $ obtained at the end of step 3:
This gradient is used to defined an updated $\bfV_{\Hxc}^{(s+1)}[\bfn]$ as 
  \begin{eqnarray}
    \bfV_{\Hxc}^{(s+1)}[\bfn]  \equiv {\rm diag}(\bmnabla[\bmg^{(s)}])  - \dfrac{\delta T_s^{(s)}[\bfn]}{\delta \bfn}.
  \end{eqnarray}
$\delta T_s^{(s)}[\bfn]/\delta \bfn$ 
can be computed using the non-interacting pKS system as 
  \begin{eqnarray}
       \dfrac{\delta T_s^{(s)}[\bfn]}{\delta \bfn} = \mu^{{\rm pKS}(s)}\cdot\mathbb{1}_1 - \bfV_{{\Hxc}}^{(s)}[\bfn]
\end{eqnarray}
where $\mathbb{1}_1$ stands for the vector of size $N$ filled with 1.
This finally leads to
    \begin{eqnarray}
  \bfV_{\Hxc}^{(s+1)}[\bfn]= \bfV_{\Hxc}^{(s)}[\bfn] +  {\rm diag}(\bmnabla[\bmg^{(s)}]) - \mu^{{\rm pKS}(s)}\cdot\mathbb{1}_1. \label{eq:V_Hxc_align}
  \end{eqnarray}
If $\bfV_{\Hxc}^{(s+1)}[\bfn] = \bfV_{\Hxc}^{(s)}[\bfn]$ convergence is reached since the interacting chemical potential $\mu$ is equal to the pKS chemical potential  $\mu = \mu^{\rm pKS}$.
Else, $s \rightarrow s + 1$ and go back to step 2.
\end{enumerate}
We stress that the 1-RDM $\bmg^{\rm pKS}$ obtained at step 2 of the SOFT-DIVA algorithm is not diagonal but idempotent and shows equivalent diagonal element as the interacting non-idempotent 1-RDM constructed in step 3 of the SOFT-DIVA algorithm.

\section{Results and discussions}
\label{sec:results}

We propose to study the performance of DIVA on the
one-dimensional (1D) Hubbard model and on a simple molecular system. 
For the 1D Hubbard model, in the spirit of Mitxelena {\it et al.}~\cite{mitxelena2017performance}, we compare
a NOF (the M\"uller functional) with a functional derived in lattice representation such as proposed by T\"ows and Pastor to approximate the correlation energy~\cite{tows2011lattice,tows2012spin,
tows2014density}. For molecular systems, only NOFs are available and we also employ the M\"uller functional.
In order to be self-contained, we provide in Appendix~\ref{app:funcs} a brief description of the functionals used in the manuscript.

\subsection{Hubbard model}

The homogeneous Hubbard model can be defined by the following Hamiltonian,
\begin{equation}
\hat{H} = -t \sum_{\langle ij \rangle, \sigma} \hat{c}_{i\sigma}^{\dagger}\hat{c}_{j\sigma} + \frac{U}{2} \sum_{i} \sum_{\sigma \sigma'} \hat{c}_{i\sigma}^{\dagger}\hat{c}_{i\sigma'}^{\dagger}\hat{c}_{i\sigma'}\hat{c}_{i\sigma}
\end{equation}
\begin{figure}
  \centering
  \resizebox{\columnwidth}{!}{
    \includegraphics[scale=1.0]{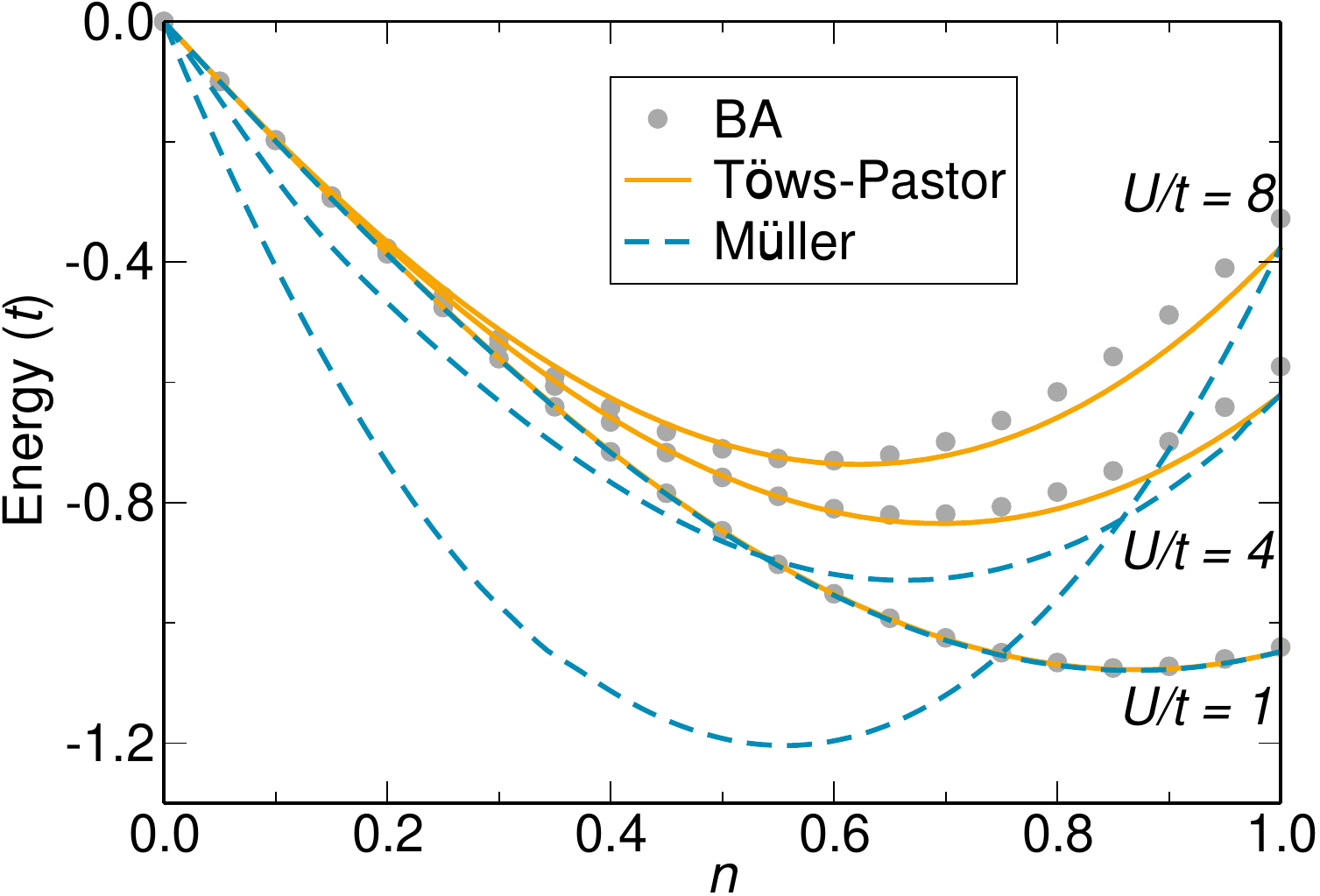}
  }   
  \caption{ Ground-state energy of the 1D Hubbard model  as a function of the electron filling $n$ at different values of the Coulomb integral $U/t$. Results are given for the exact Bethe--Ansatz (gray dots), and using DIVA with the T\"ows--Pastor functional (full orange lines) and the M\"uller functional (dashed blue lines).}
  \label{fig:Hubbard-1}
\end{figure}
where the notation $\langle \cdot \rangle$ refers to first neighbor lattice sites. 
In contrast to Eq.~(\ref{eq:H_elec}), the indices correspond to spatial orbitals (or sites) and
$\sigma = \lbrace \uparrow,\downarrow\rbrace$ denotes the spin. 
Focusing on the homogeneous model allows us to bypass
Eq.~(\ref{eq:KS_type1}) and to solely focus on the solution of Eq.~(\ref{eq:saddle_gd}) within the DIVA framework. 
In practice, chains with 202 sites have been considered with periodic boundary conditions, and the bisection algorithm has been employed for bracketing and line minimization in steps 2b and 4, respectively (see Sec.~\ref{sec:monoDIVA}).
We have considered the M\"uller functional, defined as a function of the natural orbital occupation number 
(Bl\"och function for periodic systems), 
and the T\"ows--Pastor functional that is defined in the lattice site representation. 
Note that the M\"uller functional has been investigated in the context of finite Hubbard
chains~\cite{kamil2016reduced,mitxelena2017performance}, and the T\"ows--Pastor 
functional has been applied in the context of the Anderson model~\cite{tows2011lattice,tows2012spin,tows2014density}.
In terms of  performance, we found that fast convergence is achieved with DIVA (5 iteration max to reach an energy difference of $\delta E/t < 10^{-7}$) for both functionals.
However, computational time is smaller for the T\"ows--Pastor functional compared to the M\"uller one. 
Indeed, for the M\"uller functional depending only on natural orbital occupation numbers, a standard minimization algorithm in the natural orbital representation appears well-suited~\cite{piris2009iterative}, while a minimization scheme directly in the lattice representation such as DIVA appears more adapted for the T\"ows--Pastor functional. 
For the latter, the computation of the energy and energy gradients (needed several time at each minimization step) do not require to switch between lattice and natural orbital representations.

In Fig.~\ref{fig:Hubbard-1}, the ground-state energy is shown as a function of the electron filling $n$. 
It shows that the T\"ows--Pastor functional is in fair agreement with the exact Bethe--Ansatz (BA) solution~\cite{NoMott_Hubbardmodel,shiba1972magnetic}
regardless of the electron filling and the strength of the Coulomb repulsion $U/t$.
On the contrary, the M\"uller functional leads to inaccurate energies away from the half-band filling and especially for large $U/t$.
Note that this wrong behaviour goes together with
an unphysical negative value of $\partial E /\partial U$ corresponding to the number of double occupations (not shown).
These unphysical results for the M\"uller functional arises from equivalent spin on-site exchange contributions to the energy, associated with the operator $\hat{c}_{i\sigma}^{\dagger}\hat{c}_{i\sigma}^{\dagger}\hat{c}_{i\sigma}\hat{c}_{i\sigma}$ that is supposed to be nil.

 \begin{figure}[h]
   \centering
   \resizebox{\columnwidth}{!}{\includegraphics[scale=1.0]{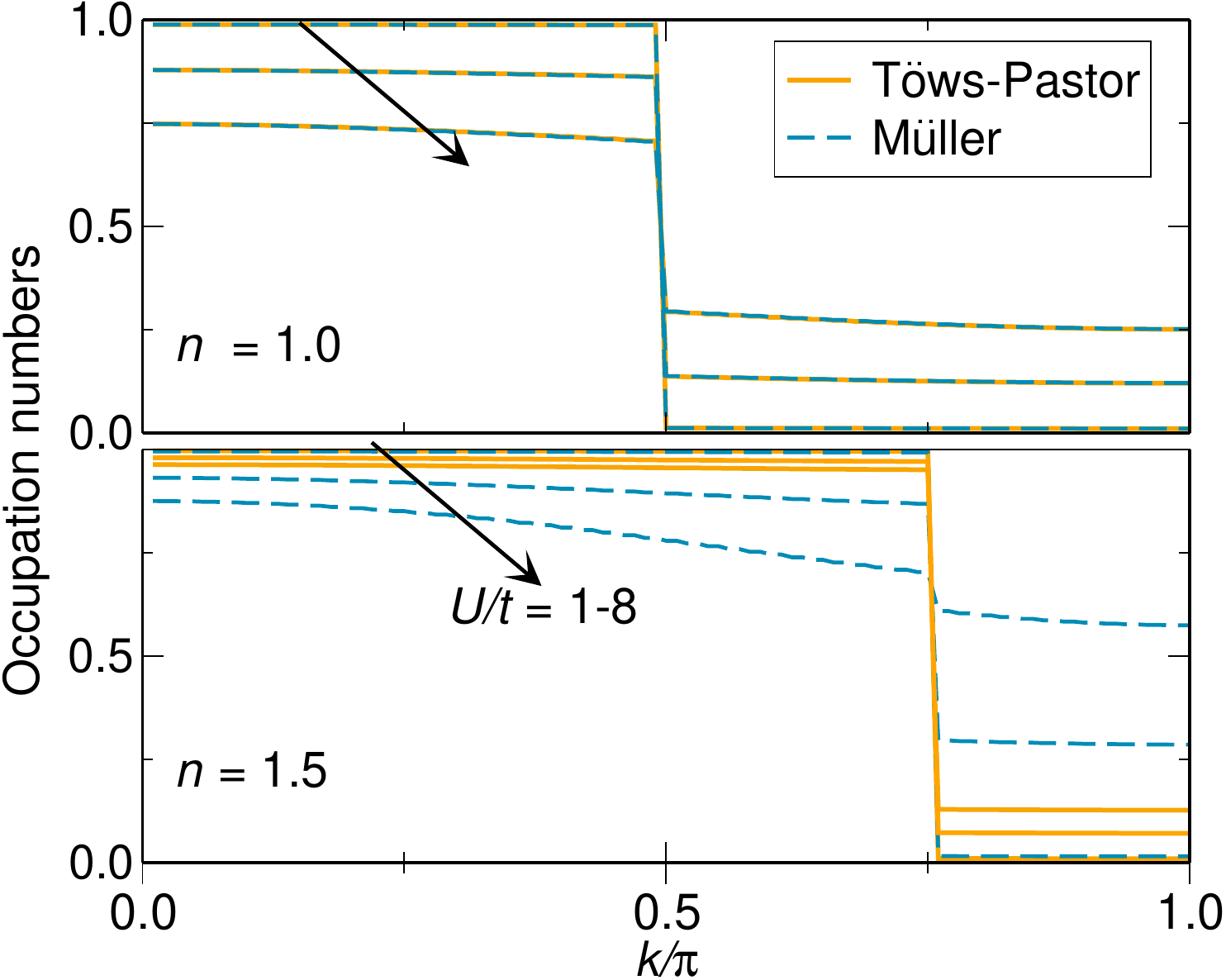}}
   \caption{ Occupation numbers as a function of the Bl\"och wave number $k$ for the half-band filled 1D Hubbard model at $U/t = 1,4$ and $8$ (in ascending order following the black arrow). 
   The top panel corresponds to the half-band filling ($n=1$) case while the bottom panel corresponds to a three-quarter filling case ($n=3/2$).
   Results are shown using the T\"ows--Pastor functional (full orange lines) and the M\"uller functional (dashed blue lines).}
   \label{fig:Hubbard_2}
 \end{figure}
 In Fig.~\ref{fig:Hubbard_2}, the occupation numbers of the natural orbitals are represented as a function of the plane-wave number $k$, at different electron fillings $n$, different strengths of the Coulomb repulsion $U/t$, and for both the T\"ows--Pastor and M\"uller functionals. 
At half-band filling ($n=1$), the T\"ows--Pastor and the M\"uller functionals lead to very similar occupation numbers with a $k$-dependency that is close to a step function, which is due the local nature of the Coulomb interaction.
The step-function-like behaviour also explains the fast convergence of the algorithm as it corresponds to a quasi homogeneous renormalization of all off-diagonal terms of the density matrix in the lattice site representation, and thus to a renormalization of the kinetic energy.
Turning to the three-quarter band filling, this behavior remains, however with large discrepancies between the T\"ows--Pastor and the M\"uller functional. 
As already mentioned away from half-band filling, the M\"uller functional leads to negative double occupations in particular when $U/t$ increases, which comes along with a drastic reduction of the kinetic energy and thus drastic modification of the occupation numbers.
In contrast, occupation numbers obtained with the T\"ows--Pastor functional are less dependent on $U/t$, which is consistent with the kinetic energy remaining finite even in the strongly interacting regime away from half-band filling.

\begin{figure}[h]
  \centering
\resizebox{\columnwidth}{!}{
  \includegraphics[scale=1.0]{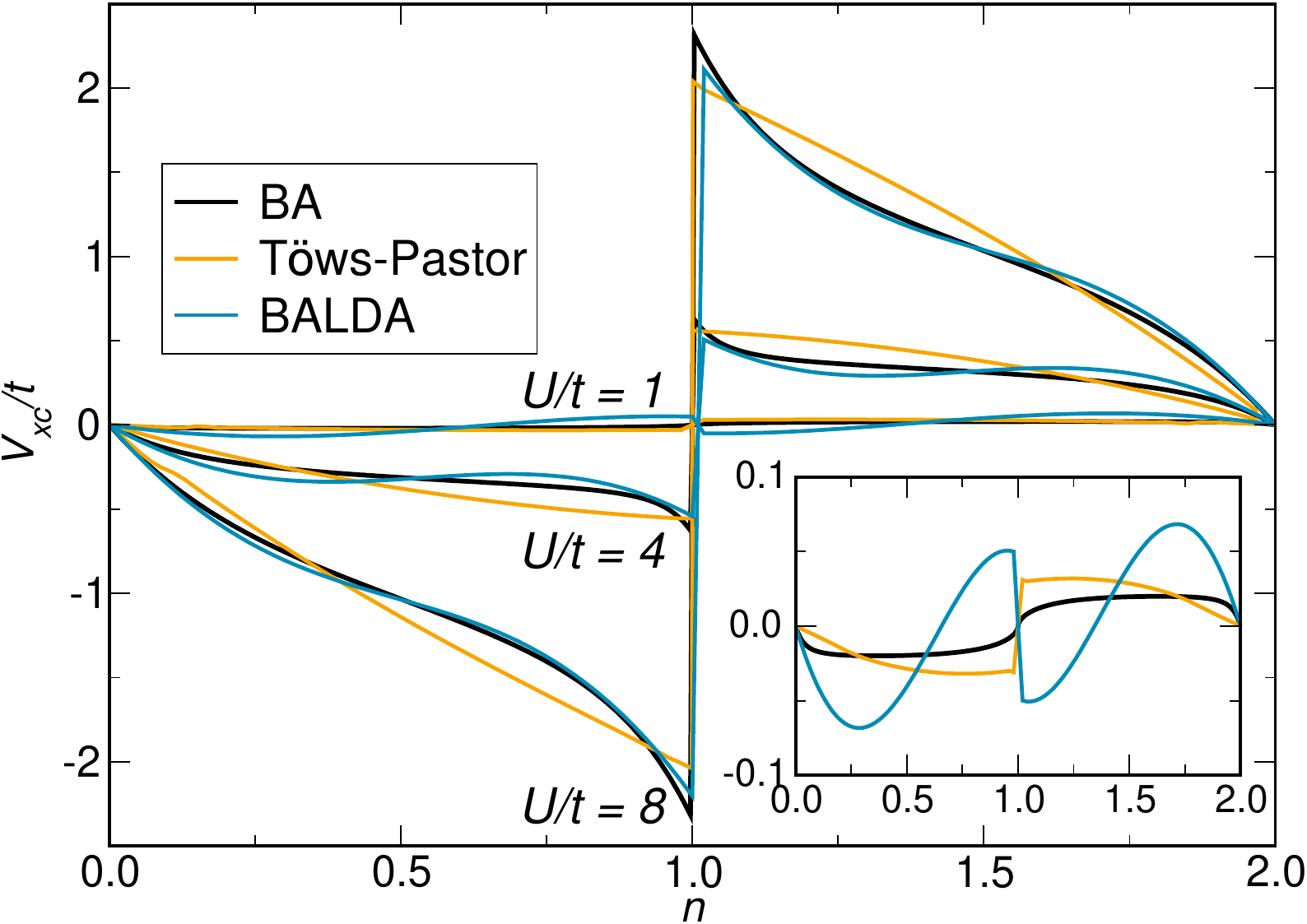}
  } 
  \caption{SOFT exchange and correlation potential $V_{xc}$ for the 1D Hubbard model as a function the electron filling $n$ and at different values of the Coulomb integral $U/t$.
Results for the T\"ows--Pastor functional are compared with BALDA and with BA.
The embedded subplot shows the curve associated to $U/t = 1$.}%
  \label{fig:Hubbard_3}
\end{figure}

Let us now investigate the per-site SOFT exchange and correlation potential $V_{xc}$ obtained by Eq.~(\ref{eq:V_Hxc_align}).
In Fig.~\ref{fig:Hubbard_3}, $V_{xc}$ obtained using the T\"ows--Pastor functional
is compared with the exact BA one
and with the Bethe--Ansatz Local Density Approximtion (BALDA)~\cite{lima2002density,lima2003density},
with respect to the electron filling $n$ and for different values of $U/t$.
$V_{xc}$ is not shown for the M\"uller functional as it leads to unphysical results away from half-band filling.
A major positive result for the T\"ows--Pastor functional is the ability to fairly reproduce the $V_{xc}$ discontinuity at half-band filling, enabling the description of the Mott--Hubbard transition from metal to insulating regimes.
This highlights the ability to design sound exchange and correlation potentials for SOFT from 1-RDMFT approximations. 
Away from half-filling ($0\leq n < 1$ and $1<n\leq 2$), the T\"ows--Pastor functional leads to a monotonous behavior of $V_{xc}$, in contrast to BALDA that shows unphysical inflection points and even positive values for $V_{xc}$, particularly in the weakly correlated regime away from the half-filled
case (see the zoomed panel in Fig.~\ref{fig:Hubbard_3}).
This unphysical behaviour was already observed
and shown in Eqs.~(31) and (32) in Ref.~\cite{senjean2018site}.
Finally, in contrast to BALDA that has been specifically designed for the 1D Hubbard model, the T\"ows--Pastor functional can \textit{a priori} be applied to any lattice dimension.

% In Fig.~\ref{fig:Hubbard_3} is shown the SOFT exchange and correlation (XC) potential $V_{xc}$ obtained through  Eq.~(\ref{eq:V_Hxc_align}) using the T\"ows--Pastor functional as a function of the electron filling $n$ and for different value of the Coulomb integral $U/t$. $V_{xc}$ is not shown  for the M\"uller functional since it leads to unphysical results away from half-band filling.  
% $V_{xc}$ obtained using the  T\"ows--Pastor functional are compared with numerically exact potential obtained using the Bethe-Ansatz and with the BALDA approximation for SOFT~\cite{limaDensityFunctionalsNot2003a}. A major positive results for the T\"ows--Pastor functional is the ability to reproduce fairly the $V_{xc}$ discontinuity at half-band filling enabling the description of Mott insulating regime. This highlight the ability of designing sound exchange and-correlation potential for SOFT from rDMFT approximation. Away from half-fillng ($0\leq n < 1$ and $1<n\leq 2$, the  T\"ows--Pastor functional leads to monotoneous behavior of $V_{xc}$ in contrast with the BALDA approximation that shown unphysical inflection point and even positive value for $V_{xc}$ in particular in the low interacting regime. Finally, in contrast to the BALDA approximation that has been specifically designed  for the 1D Hubbard model, the T\"ows--Pastor functional can be {\it a priori} applied to any lattice dimention.

\subsection{Molecular sytems}

In contrast to the Hubbard model,
both Eqs.~(\ref{eq:KS_type1}) and (\ref{eq:saddle_gd}) have to be solved
for molecular systems.
As a proof of concept, we apply our three different DIVA implementations described in Sec.~\ref{sec:implementations} to the hydrogen molecule H$_2$ in the 6-31G basis,
using the M\"uller functional.
As DIVA is general and not tight to any particular orbital representation, we decided to work in the orthogonal atomic orbital (OAO) basis whose first orbital occupations are given by a 
DFT calculation using a local spin density approximation functional (SVWN) from the PSI4 program~\cite{psi4}.
Numerical energy gradients with respect to $\bmg^{\neq}$ and $\bfn$ have been computed with finite difference.% with $\epsilon = 10^{-6}$.

\subsubsection{Convergence of the mono-parameter, multi-parameter, and SOFT-DIVA}

\begin{figure}
  \centering
\resizebox{\columnwidth}{!}{
  \includegraphics[scale=1.0]{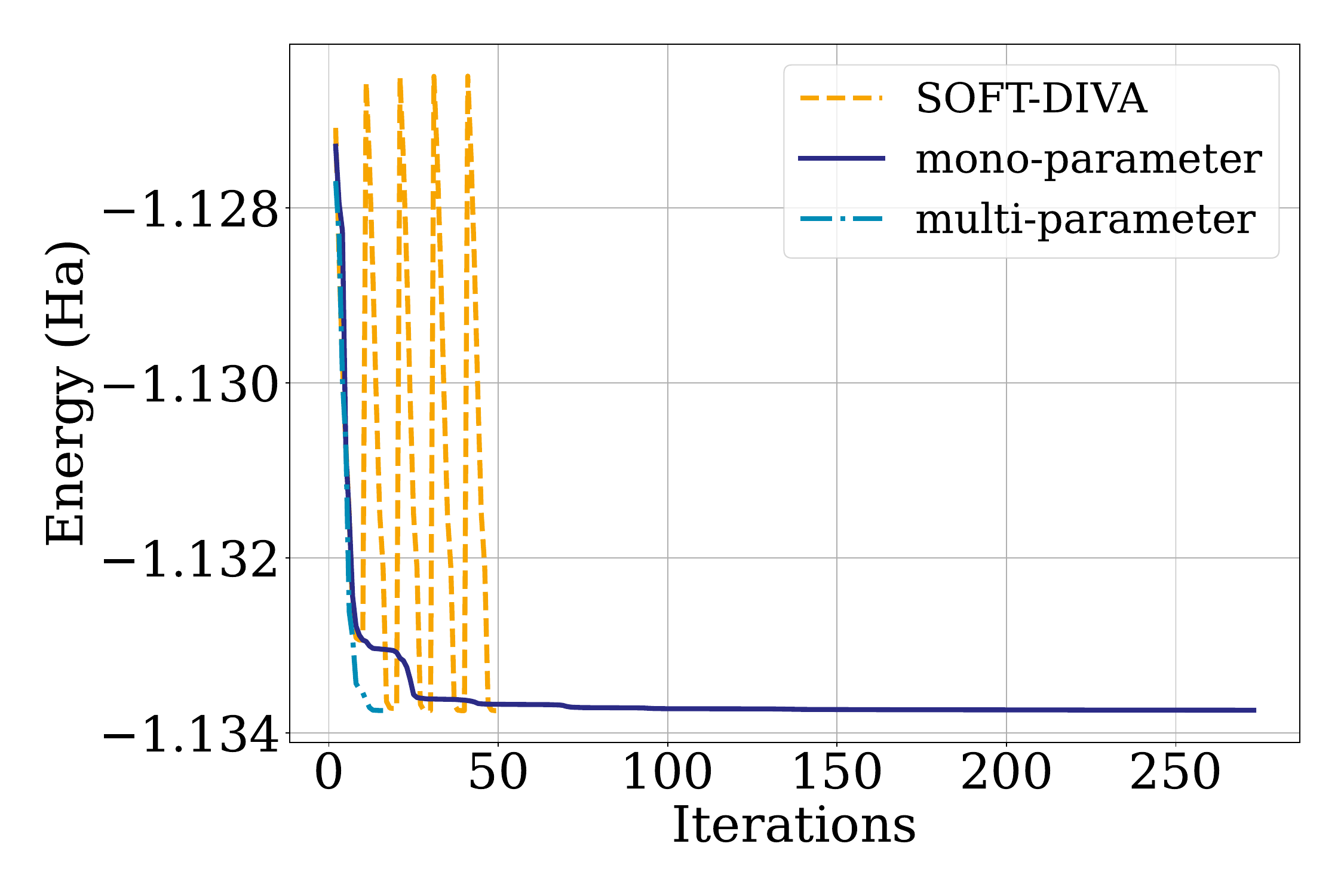}
  }
  \caption{1-RDMFT energy with respect to the number of iterations, using SOFT-DIVA, mono-parameter DIVA, and multi-parameter DIVA, for $R=1.0$~{\AA}. The 6-31G basis and the M\"uller functional were used.}%
  \label{fig:convergence}
\end{figure}

In Fig.~\ref{fig:convergence}, we compare the convergence of the energy with respect
to the number of iterations, using either the mono- or multi-parameter DIVA implementations, for an interatomic distance of $R=1.0$~\AA.
A convergence criteria of $10^{-8}$ Hartree for the energy and
$10^{-5}$ for the Frobenius norm of the 1-RDM have been used.
As expected, the multi-parameter implementation converges in much fewer iterations than the mono-parameter one, although it does not mean that it is
computationally faster than the mono-parameter DIVA.
Indeed, each iteration of mono-parameter DIVA has the same cost, i.e. the cost of the optimization of a single variational parameter $z$.
On the contrary, the number of parameters to optimize in multi-parameter DIVA increases at each additional iteration.
Hence, there is a trade-off between the number of iterations and the cost of optimizing many parameters at once.
Note that in this work, multi-parameter DIVA was always computationally much faster than mono-parameter DIVA.
Besides, the former always converges in around 20 iterations
while thousands of iterations are sometimes required using mono-parameter DIVA for some interatomic distances.
This very slow convergence has also been observed in other gradient-based algorithms and is attributed to the fact that NOFs are not proper functionals of the 1RDM, since these functionals are not invariant with respect to the rotation of two orbitals having degenerate occupation numbers~\cite{cancesProjectedGradientAlgorithms2008}.
We believe that the convergence of mono-parameter DIVA could be significantly improved by using analytical gradients.
Alternatively, considering a simultaneous perturbation stochastic approximation
can reduce significantly the number of energy evaluations needed to compute the numerical gradient, compared to standard finite difference.
This is left for future work.

Let us now consider the convergence of SOFT-DIVA,
where the multi-parameter implementation has been used for the optimization of the off-diagonal elements of the 1-RDM [Eq.~(\ref{eq:saddle_gd})].
As readily seen in Fig.~\ref{fig:convergence}, there is no numerical advantage in splitting the variational principle into two parts.
Indeed, although constraining the orbital occupations fixed leads to faster convergence
when solving Eq.~(\ref{eq:saddle_gd}), one has to do it for each new orbital occupations obtained when solving Eq.~(\ref{eq:KS_type1}) with the previously obtained Hxc potential, which is manifested
by each peak of SOFT-DIVA in Fig.~\ref{fig:convergence}.
However, in practice it allows for the estimation of the Hxc potential, functional of the orbital occupation, at the M\"uller functional level of approximation.

\subsubsection{Energies and occupation numbers}

\begin{figure}
  \centering
  \resizebox{\columnwidth}{!}{ \includegraphics[width=\textwidth]{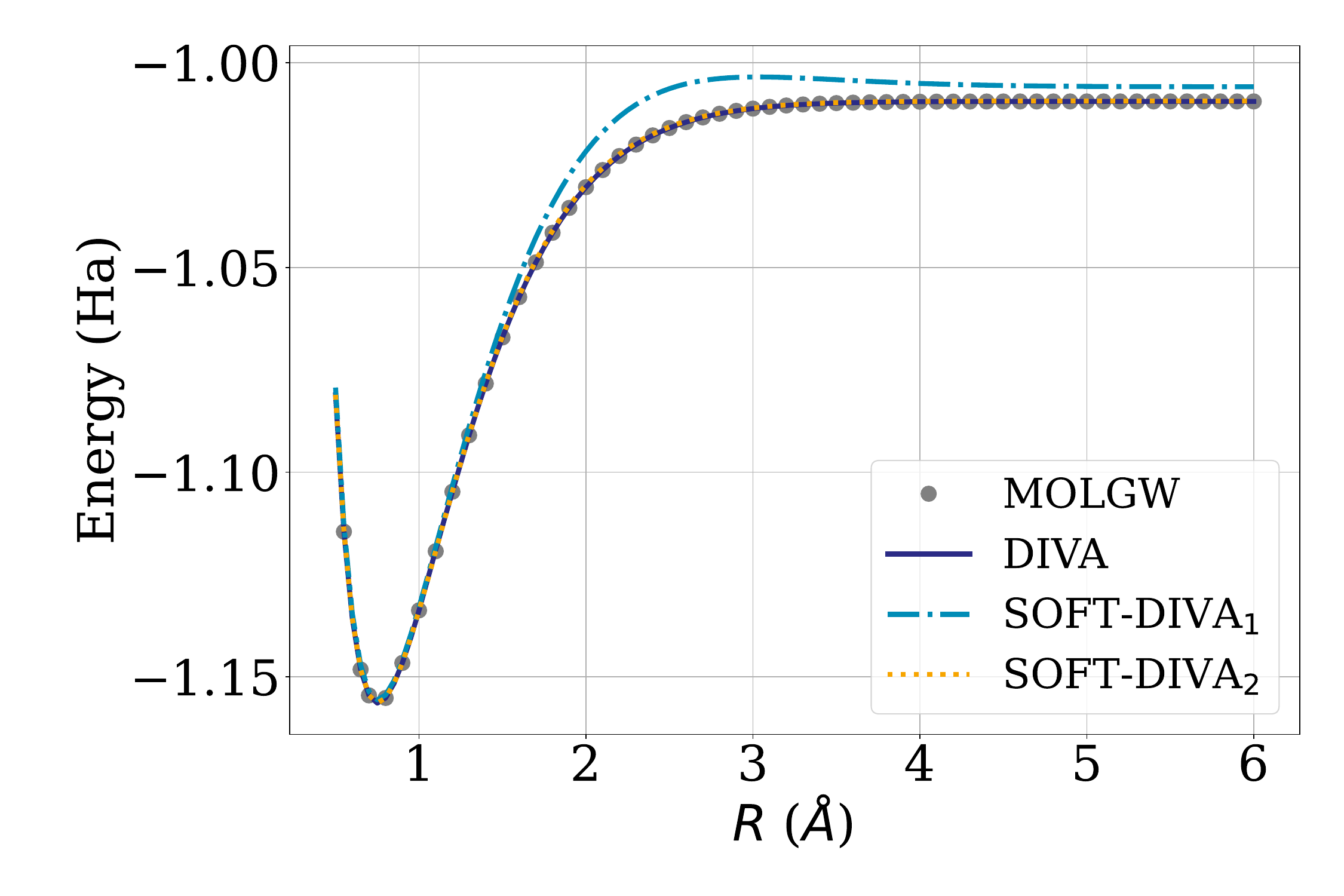}}
    \caption{1-RDMFT energy with respect to the interatomic distance in the 6-31G basis using the M\"uller functional.
SOFT-DIVA$_n$ corresponds to the $n$-th iteration of SOFT-DIVA, i.e. when the off-diagonal elements have been optimized for a fixed orbital occupation.}
\label{fig:energy} 
\end{figure}

\begin{figure}
  \centering
  \resizebox{\columnwidth}{!}{      \includegraphics[width=\textwidth]{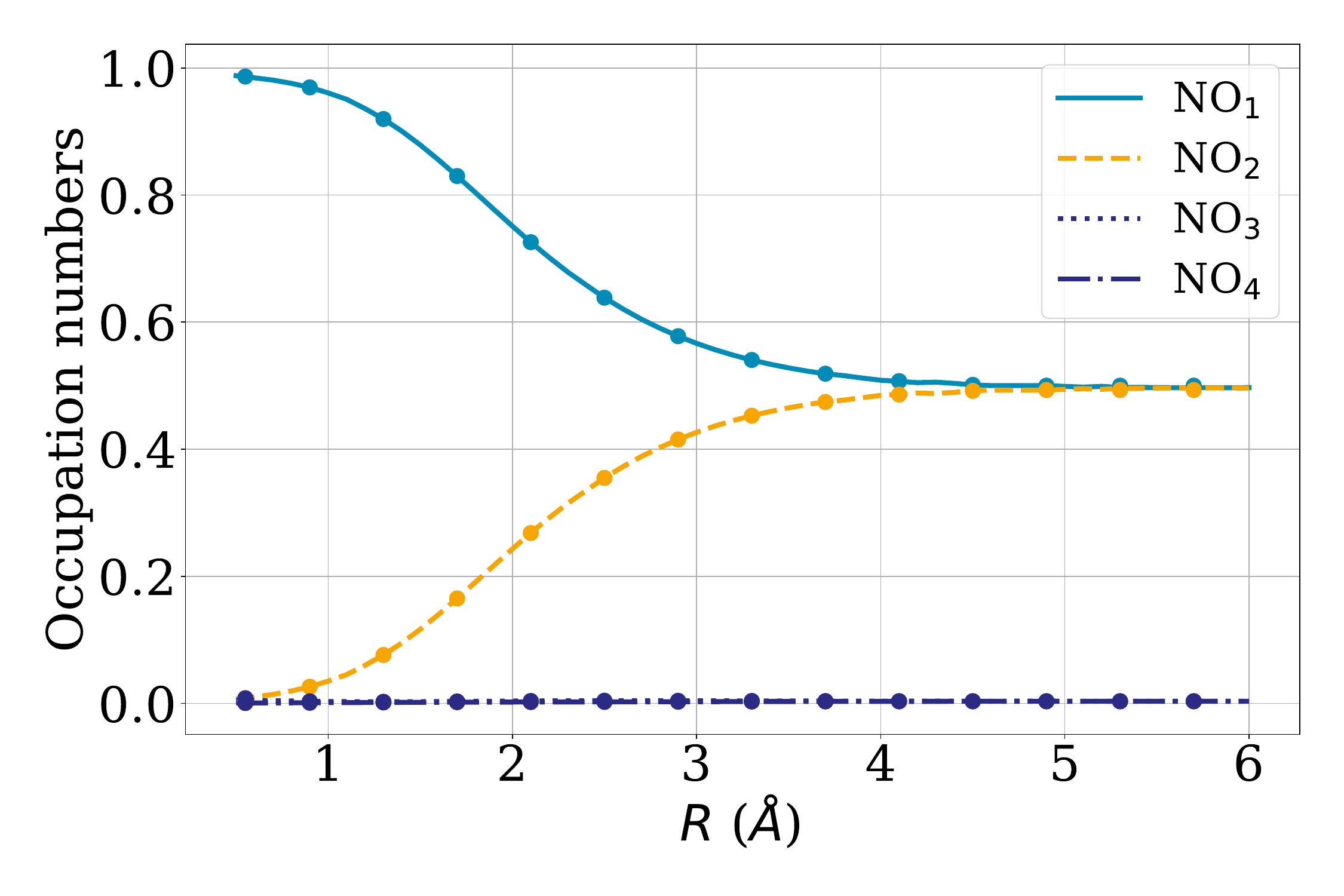}}
    \caption{Occupation numbers of each natural orbital $i$ denoted NO$_i$ with respect to the interatomic distance. Dots correspond to the reference results obtained with MOLGW.}
    \label{fig:occupation}
\end{figure}

In order to check the accuracy of our implementation, we compare the energy and occupation numbers obtained by multi-parameter DIVA with the 1-RDMFT implemented in MOLGW~\cite{bruneval2016molgw,mauricio_molgw}.
As readily seen in Figs.~\ref{fig:energy}
and \ref{fig:occupation},
the multi-parameter DIVA energy and occupation numbers are on top of the reference ones obtained with MOLGW.
One can also see how the fractional occupation numbers of the natural orbitals in Fig.~\ref{fig:occupation} are slowly getting degenerate when stretching the bond,
which is a very well known result showing that 1-RDMFT can easily account for static correlation, in contrast to KS-DFT for which the 1-RDM is idempotent by construction.

Turning to SOFT-DIVA, by stopping the algorithm at each new update of the Hxc potential, one can see how much the change in orbital occupations impacts the energy.
In other words, one can look at the best
1-RDM functional energy corresponding to a fixed orbital occupation (in the OAO basis), where only the off-diagonal elements of
the 1-RDM have been optimized.
In Fig.~\ref{fig:energy}, 
the energy of the first iteration deviates from the
reference one when stretching the hydrogen bond,
meaning that the orbital occupations
are not optimal in this case.
This deviation is already corrected
at the second iteration, corresponding to the first update of the Hxc potential that gives sufficiently accurate orbital occupations for the energy to be on top of the converged one.

\begin{figure}
  \centering
  \resizebox{\columnwidth}{!}{\includegraphics[width=\textwidth]{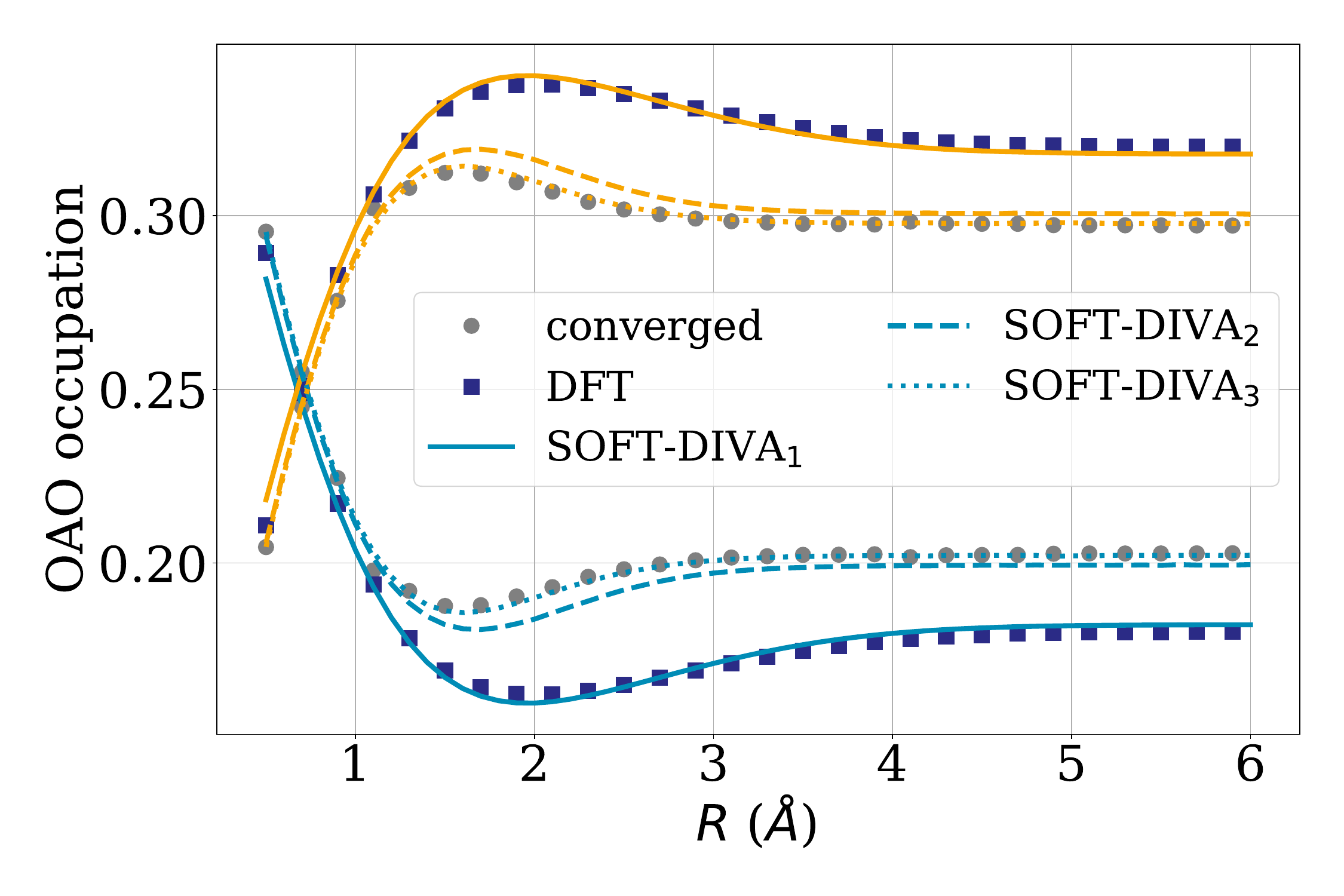}}
    \caption{OAO occupations with respect to the interatomic distance.
    For symmetry reasons, there are only two different OAO occupation values (corresponding to the 1s and 2s of the hydrogens), shown in blue and orange, respectively. Converged DIVA occupations and DFT ones are depicted with circles and squares, respectively. SOFT-DIVA$_n$ corresponds to the $n$-th iteration of SOFT-DIVA.}
\label{fig:OAO} 
\end{figure}

Turning to the orbital occupations,
the occupations
of the OAO are given for the first three iterations of SOFT-DIVA in Fig.~\ref{fig:OAO}.
Note that the initial orbital occupations of SOFT-DIVA are obtained by taking the diagonal (in the OAO basis) of the Hxc potential from the converged DFT calculation, which explains why they are not equal but close to the DFT occupations.
According to Fig.~\ref{fig:OAO}, the OAO occupations at the second iteration of SOFT-DIVA are indeed much closer to the converged one, 
thus explaining why the energy in Fig.~\ref{fig:energy} matches the converged one.
However, another iteration is required to get very accurate OAO occupations.

\section{Conclusions and perspectives}
\label{sec:conclu}

We have introduced a new variational minimization scheme for 1-RDMFT called density-matrix interpolation variational ansatz (DIVA) that works in any orbital basis representation while satisfying the ensemble $N$-representability conditions.
We have derived different implementations 
and successfully tested them on the uniform Hubbard model with
the M\"uller and the T\"ows--Pastor functionals,
as well as the hydrogen molecule with the M\"uller functional.
In summary,
every implementations give the correct 1-RDM and energy.
The mono-parameter DIVA converges in a few iterations for the Hubbard model, but exhibits very slow convergence for the hydrogen molecule.
The multi-parameter DIVA led to a much faster convergence,
at the expense of more involved optimizations of the variational parameters.
Finally, by allowing the estimation of the
exchange and correlation potential for a given orbital occupation vector, the SOFT-DIVA implementation contributes to the development of orbital occupation functionals for chemistry.
Besides, we stress that DIVA is more adapted to the computational basis representation, as it would require less swapping between the computational and the NO basis for the calculation of energy and energy gradients. 
As also mentioned by Canc\`es and Pernal~\cite{cancesProjectedGradientAlgorithms2008} such type of algorithm is expected to perform efficiently 
for proper functionals of the 1RDM, that can be expressed in any orthonormal orbital
basis. 
To date, only the Hartree--Fock functional fulfills this property. 
Hence, this work motivates the future development of proper functionals of the 1RDM, beyond NOFs.
The implementation of analytical gradients will also be
essential for targeting 1RDMs that are too close to the boundary, where numerical gradients are ill-defined.
Work is in progress in these directions.

\section{Acknowledgments}

We thank Mauricio Rodriguez-Mayorga for fruitful discussions about 1-RDMFT in general.
QM and MS would like to thank the ANR (Grant No. ANR-19-CE29-0002 DESCARTES project) for funding.

\appendix

\section{Convexity of the ensemble $N$-representable 1-RDM domain.}
\label{App:demo-convex}
The  ensemble $N$-representable domaine $\Omega_{\bmg}$ is convex if, for 
any ensemble $N$-representable 1-RDM $\bmg^A$ and $\bmg^B$,
$\bmg  = z\bmg^A + (1-z)\bmg^B$ with $z \in [0,1]$ is also  ensemble $N$-representable. \\ 
Let us consider a $N$ dimensional vector $\bfx$. Thus we have 
\begin{equation}
\bfx^T \bmg \bfx = z \bfx^T \bmg^A  \bfx  + (1-z) \bfx^T  \bmg^B  \bfx  \geq 0
\end{equation}
since $\bfx^T \bmg^A  \bfx \geq 0 $ and  $\bfx^T  \bmg^B  \bfx \geq 0$ ($\bmg^A$ and $\bmg^B$ are positive semi-definite) and $0 \leq z \leq 1$.
Consequently, $\bmg$ is also positive semi-definite. Similarly,
\begin{equation}
\bfx^T \left( \mathbb{1} - \bmg \right)\bfx = z \bfx^T \left( \mathbb{1} - \bmg^A \right) \bfx  + (1-z) \bfx^T \left( \mathbb{1} - \bmg^B \right) \bfx  \geq 0
\end{equation}
since  $0 \leq z \leq 1$ and $\bfx^T \left( \mathbb{1} - \bmg^A \right)   \bfx \geq 0 $ and $\bfx^T  \left( \mathbb{1} - \bmg^B \right)\bfx \geq 0$) since $\mathbb{1} - \bmg^A$ and $\mathbb{1} - \bmg^B$ are positive semi-definite. Consequently,  $\mathbb{1} - \bmg$ is also semi positive definite. Finally, $\bmg$ and  $\mathbb{1} - \bmg$ being  positive semi-definite, $\bmg$ is  ensemble $N$-representable  and $\Omega_{\bmg}$ is convex. 

\section{Interior and boundary of $\Omega_{\gamma}$ }
\label{App:demo-topo}
 We define the neighborhood $\mathcal{N}(\bmg)$ of a matrix $\bmg \in \mathcal{M}_N(\mathbb{R})$ as  the open ball
%\begin{align}
 %\mathcal{N}(m) = \left\lbrace m' \; | \; m' = m + \epsilon  m'',\; m'' \in \mathcal{M}_N(\mathbb{R}) \; \& \; \epsilon \rightarrow 0 \right\rbrace, 
 %\end{align}
 \begin{align}
 \label{eq:open_ball2}
 \mathcal{N}(\bmg) = \left\{ \bmg'  \in  \mathcal{M}_N(\mathbb{R}) \; | \; || \bmg - \bmg'  || < \epsilon, \epsilon \rightarrow 0 \right\} 
 \end{align}
where $||\cdot||$ refers to the Frobenius norm. We consider $\bmg' \in  \mathcal{N}(\bmg)$ and $\delta \bmg =  \left(\bmg'  - \bmg \right)/\lambda $ with $\lambda=  ||\bmg'  - \bmg||< \epsilon$, such that 
\begin{align}
\label{eq:gamma_vois}
  \bmg' = \bmg + \lambda \delta\bmg.
 \end{align}
 Following a perturbation approach with $\bmg' = \bmg + \lambda \delta \bmg$ and $\lambda \rightarrow 0$, we propose to develop the eigenvalues $\bmn' = \bmn + \lambda \delta \bmn$ and eigenvectors $\bfu' = \bfu + \lambda \delta \bfu $ as a power of infinitesimal $\lambda$ where the $\bmn$ and $\bfu$ correspond to the eigenvalues diagonal matrix and eigenvector matrix of $\bmg$, respectively, 
 \begin{equation}
 \label{eq:eta_gamma0}
  \bmg \bfu^{\dagger} =\bfu^{\dagger} \bmn.
 \end{equation}
The unitary transformation $\bfu'$ that diagonalizes $\bmg'$ corresponds to 
\begin{equation}
  \bmg' \bfu'^{\dagger} = \bfu'^{\dagger} \bmn',
\end{equation}
and Eq.~(\ref{eq:eta_gamma0}) is recovered at the zero order development. 
At first order it reads
\begin{equation}
  \delta \bmg \bfu^{\dagger} +  \bmg \delta \bfu^{\dagger} - \delta \bfu^{\dagger} \bmn - \bfu^{\dagger} \delta \bmn = 0.
\end{equation}
Multiplying by $\bfu$ on the left side leads to
\begin{eqnarray}
  & & \bfu \delta \bmg \bfu^{\dagger} +  \bfu \bmg {\delta \bfu }^{\dagger} - \bfu{ \delta \bfu}^{\dagger}\bmn - \bfu \bfu^{\dagger} \delta \bmn = 0 \nonumber \\
  & &  \bfu \delta \bmg {\bfu}^{\dagger}  -  \delta\bmn + \left[\bmn, \bfu {\delta \bfu }^{\dagger}\right] = 0. \label{eq:pert_rep}
\end{eqnarray}
On the diagonal part of Eq.~(\ref{eq:pert_rep}), the last term on the right hand side cancels, such that
\begin{equation}
  \delta \bmn = \bfu \delta \bmg {\bfu}^{\dagger}.  
\end{equation}
Let us define 
\begin{align}
d(\bmg) = {\rm min}\left(\eta_1,  1 - \eta_N \right),
\end{align}
where the eigenvalues are sorted in ascending order.
$d(\bmg)$ appears as a pseudo-distance from $\delta \Omega_\bmg$ as $d(\bmg) \geq 0$ for $\bmg \in \Omega_\bmg$, $d(\bmg) > 0$ for $\bmg \in \mathring{\Omega}_{\bmg}$, $d(\bmg) = 0$ for $\bmg \in \delta \Omega_\bmg$ and finally, $d(\bmg) < 0$ for non-representable matrices $ \bmg \notin  \Omega_\bmg$.  
It follows that
\begin{align}
d(\bmg') = {\rm min}\left(\eta_1 +  \delta \eta_1,  1 - (\eta_N + \delta \eta_N) \right),
\end{align}
and $ d(\bmg')$ is contained in the interval $ d(\bmg')$ $ \in$ $] d(\bmg) - \lambda,d(\bmg) + \lambda [ $ since $ |\delta \bmn | \leq  \lambda |\bfu  \delta \bmg  \bfu^\dagger| \leq \lambda $.

Consequently, for each $\bmg  \in \mathring{\Omega}_{\bmg}$ there is a neighborhood with $\epsilon < d(\bmg)$ such that $\mathcal{N}(\bmg) \subseteq \Omega_{\bmg}$, thus making $\mathring{\Omega}_{\bmg}$ the topological interior of $ \Omega_{\bmg}$. 
On the contrary, for $\bmg  \in \delta \Omega_{\bmg}$, since $d(\bmg) = 0$ it becomes trivial to find matrices in $ \mathcal{N}(\bmg) $ that do not belong to $\Omega_{\bmg}$. Hence, $\delta \Omega_{\bmg}$ is the topological boundary of $\Omega_{\bmg}$. 
 
\section{Density-matrix decomposition }

\label{App:demo-decomp}

We show that for any non-idempotent $\bmg$, there exists at least one decomposition in terms of (up to $2^N$) idempotent matrices such that
 $\bmg = \sum_s z_s \bmg_*^{(s)}$ with $\sum_s z_s = 1$. 
 To that aim, we propose the following iterative algorithm.
 Consider $\bmg \in \Omega_{\bmg} \backslash  \delta \Omega_{\bmg}^*$ that has $F(\bmg)$ fractional eigenvalues.
 
 \begin{enumerate}
 \item Choose one among the $F(\bmg)$ fractional eigenvalues $\eta_{k}$
 \item Construct, in the diagonal representation of $\bmg$,
 \begin{align}
  &\bmg'(\lambda) =  \bmg + \lambda \bmg^{(k)} \\
 &\gamma^{(k)}_{ij} = 0 \, \forall \,  (i, j) \neq (k,k), \,{\rm and}\,  \gamma^{(k)}_{kk} =  1
\end{align}
 with $\lambda$ a scalar
 and $\bmg'(\lambda) \in \Omega_{\bmg}$ for $\lambda \in [-\lambda^A, \lambda^B ]$ with $\lambda^A =  \eta_{k}$ and  $\lambda^B =   1 - \eta_{k}$. 
 By construction, $F\left(\bmg'(-\lambda^A)\right) = F\left(\bmg'(\lambda^B)\right) =  F(\bmg) -1$.
 \item Decompose $\bmg$ in terms of matrices with higher number of integer eigenvalues,
\begin{align}
\label{eq:gamma_decomp}
\bmg = (1- \eta_k) \bmg'(-\lambda^A) + \eta_k\bmg'(\lambda^B)
\end{align}
%with $z = \lambda^A/(\lambda^B-\lambda^A)$.
 \end{enumerate}

Repeat steps 1 and 2 for $\bmg'(-\lambda^A)$ and $\bmg'(\lambda^B)$ and insert their decomposition into Eq.~(\ref{eq:gamma_decomp}). 
Continue the algorithm until 
\begin{align}
\label{eq:gamma_decomp2}
\bmg =  \sum_s z_s \bmg_*^{(s)},
\end{align} 
with $\bmg_*^{(s)}  \in \delta \Omega_\bmg^*$ and $\sum_s z_s = 1$. 
Each iteration of the algorithm multiplies by two the number of matrices in the decomposition, making the sum in Eq.~(\ref{eq:gamma_decomp2}) running ultimately up to $2^{F(\bmg)}$ elements. 

\section{Functionals}
\label{app:funcs}
 
NOF approximations rely on the Hartree--Fock form of the two-particle reduced density matrix (2-RDM) $\Gamma_{ij,kl}^{\sigma,\sigma'}$, that can be expressed in terms of occupation numbers of the 1-RDM in the natural orbital representation as
  \begin{align}
    & \Gamma_{ij,kl}^{\sigma,\sigma} = \frac{\eta_i\eta_j}{2}\left(\delta_{ik}\delta_{jl} - \delta_{il}\delta_{jk}\right), \label{eq:G_HF1}\\
    & \Gamma_{ij,kl}^{\sigma,\bar{\sigma}} = \frac{\eta_i\eta_j}{2}\delta_{ik}\delta_{jl}.\label{eq:G_HF2}
\end{align}
A common recipe to build ersatz form of the interacting 2-RDM consists in extrapolating the same-spin elements of Eq.~(\ref{eq:G_HF1}) as
  \begin{align}
     \Gamma_{ij,kl}^{\sigma,\sigma} = \frac{\eta_i\eta_j}{2}\delta_{ik}\delta_{jl} - \frac{\mathcal{F}(\eta_i,\eta_j)}{2}\delta_{il}\delta_{jk}, \label{eq:G_NOF}
\end{align}
where the function $\mathcal{F}(\eta_i,\eta_j)$ differs from one approximation to another.
For the M\"uller functional,
it is given by~\cite{rodriguez2017comprehensive,muller1984explicit}
\begin{align}
    \mathcal{F}(\eta_i,\eta_j) = \sqrt{\eta_i \eta_j}. \label{eq:F_Muller}
\end{align}
Note that such direct extrapolation for the 2-RDM in NOF leads to violation
of the representability conditions of the 2-RDM, see for instance Ref.~\cite{rodriguez2017comprehensive} and reference therein.
%
% To avoid part of the 2-RDM representability issues, PNOF have been constructed by using some of the 2-RDM $N$-representability conditions~\cite{piris2006new,piris2013natural,piris2014perspective,rodriguez2017comprehensive}. Within the PNOF framework, the 2-RDM takes the following form
% \begin{align}
%      & \Gamma_{ij,kl}^{\sigma,\sigma} = \frac{\eta_i\eta_j -\Delta_{ij}}{2}\left(\delta_{ik}\delta_{jl} - \delta_{il}\delta_{jk}\right), \label{eq:G_PNOF1}\\
%     & \Gamma_{ij,kl}^{\sigma,\bar{\sigma}} = \frac{\eta_i\eta_j-\Delta_{ij}}{2}\delta_{ik}\delta_{jl} + \frac{\Pi_{ik}}{2}\delta_{ij}\delta_{kl}, \label{eq:G_PNOF2}
% \end{align}
% where $\eta_i$ still stands for occupation numbers of the 1-RDM (in the natural orbital reprersentation). Different expression for the matrices $\boldsymbol{\Delta}$ and $\boldsymbol{\Pi}$ characterize the different implementations of the PNOFi (i = 1–7), see Ref.~\cite{rodriguez2017comprehensive} for the specification of each PNOF. For instance within the PNOF5, $\Delta_{ij} = \eta_i \eta_j$ and $\Pi_{ij} = -\sqrt{\eta_i \eta_j}
% $ if at least one of the natural orbital $i$ or $j$ is occupied and $ \Delta_{ij} = \eta_i \eta_j$ and $\Pi_{ij} = \sqrt{\eta_i \eta_j}$ if both the natural orbital $i$ and $j$ are empty~\cite{piris2011natural}.

For the specific case of the Hubbard model, we have also used the T\"ows-Pastor functional that consists in an approximation of the on-site and opposite spin elements of the 2-RDM $\Gamma_{ii,ii}^{\sigma,\bar{\sigma}}$~\cite{tows2011lattice,tows2012spin,tows2014density}. The form of the latter is extracted from a two-level system with renormalized interactions. 
More precisely,
\begin{equation*}
      \Gamma_{ii,ii}^{\sigma,\bar{\sigma}} = \left\{
    \begin{array}{ll}
         n_i - \frac{g^2}{m - 2\sqrt{g_0^2 - g^2}} & \mbox{ if } g^2 > g_{\infty}^2 \\
      0 & \mbox{if } g^2 \leq g_{\infty}^2, ~M \leq 1 \mbox{ and } \eta_i\leq M \\
      2n_i - M  & \mbox{if } g^2 \leq g_{\infty}^2,~ M \leq 1 \mbox{ and } \eta_i > M \\
      2n_i -1  & \mbox{if } g^2 \leq g_{\infty}^2,~ M > 1 \mbox{ and } \eta_i > M \\
      M -1  & \mbox{if } g^2 \leq g_{\infty}^2,~ M > 1 \mbox{ and } \eta_i > M 
    \end{array}
                 \right.,
\end{equation*}
with $g^2 = \sum_{j \neq i} \gamma_{ij}^2$, $m = \min{(M, 2-M)}$, $M = (n_i + n')$, $n'=\sum_{kl}\gamma_{ik}\gamma_{il}\gamma_{kl}/g^2$,
\begin{equation*}
     g_0^2 = \left\{
    \begin{array}{ll}
         4 n_in'& \mbox{if } M \leq 1 \\
      4(1-n_i)(1-n') & \mbox{if } M > 1
    \end{array}
                 \right.
\end{equation*}
and finally,
\begin{equation*}
     g_{\infty}^2 = \left\{
    \begin{array}{ll}
      4 n_i(n' - n_i) & \mbox{if } M \leq 1 \mbox{ and } n_i \leq M \\
      4 n'(n_i - n')   & \mbox{if } M \leq 1 \mbox{ and } n_i > M \\
      4 (1-n_i)(n_i - n')   & \mbox{if } M > 1  \mbox{ and }n_i > M \\
      4 (1 -n')(n' - n_i) & \mbox{if } M > 1 \mbox{ and } n_i \leq M. \\
    \end{array}
                 \right.
\end{equation*}

%\bibliography{../../../../biblio}
%merlin.mbs apsrev4-1.bst 2010-07-25 4.21a (PWD, AO, DPC) hacked
%Control: key (0)
%Control: author (8) initials jnrlst
%Control: editor formatted (1) identically to author
%Control: production of article title (-1) disabled
%Control: page (0) single
%Control: year (1) truncated
%Control: production of eprint (0) enabled
\newcommand{\Aa}[0]{Aa}

\end{document}